\documentclass[12pt]{article}
\usepackage{amsmath}
\usepackage{graphicx,psfrag,epsf}
\usepackage{enumerate}
\usepackage{psfrag,epsf}
\usepackage{url} 
\usepackage{amsmath,amssymb,amsfonts,amsthm,mathtools,mathrsfs}

\usepackage{graphicx}
\usepackage{caption}
\usepackage{subcaption}
\usepackage{booktabs}

\usepackage{booktabs}
\usepackage[table]{xcolor}
\usepackage{bm,bbm}
\usepackage{color}

\usepackage{algorithm,algpseudocode}
\usepackage[symbol]{footmisc}
\usepackage{scalefnt}
\usepackage{authblk} 
\usepackage{multirow,centernot}
\usepackage[colorlinks=true, citecolor=blue, urlcolor=blue]{hyperref}
\usepackage{booktabs}
\usepackage[table]{xcolor}

\usepackage{epstopdf} 

\graphicspath{{./}{figures/}}
\DeclareGraphicsExtensions{.pdf,.eps,.png,.jpg}

\usepackage{natbib}
\usepackage{dsfont}
\usepackage[title]{appendix}
\input cyracc.def

\usepackage[left=1in,top=1.1in,right=0.5in,bottom=1in]{geometry}

\theoremstyle{definition}

\newtheorem{theorem}{Theorem}[section]

\newtheorem{corollary}[theorem]{Corollary}
\newtheorem{example}{Example}[section]

\newtheorem{lemma}[theorem]{Lemma}

\newtheorem{remark}[theorem]{Remark}

\makeatletter
\def\@seccntformat#1{\@ifundefined{#1@cntformat}%
	{\csname the#1\endcsname\quad}
	{\csname #1@cntformat\endcsname}
}
\makeatother

\newif\ifShowComments
\ShowCommentstrue
\def\strutdepth{\dp\strutbox}
\def\druk#1{\strut\vadjust{\kern-\strutdepth
        {\vtop to \strutdepth{%
                \baselineskip\strutdepth\vss
                        \llap{\hbox{#1}\quad}\null}}}}

\title{\bf
A transformed-score approach to closed-form and one-step efficient estimation for the beta distribution  
} 

\author{
\text{Roberto Vila}$^{1}$,
\text{Helton Saulo}$^{1,2}$\thanks{{\bf Corresponding author:} Helton Saulo, email: {heltonsaulo@gmail.com}
	\newline
}\,\,,
\text{Felipe Quintino}$^{1}$\,\,and
\text{Pedro~Luiz~Ramos}$^{3}$\\
{\small $^{1}$Department of Statistics, University of Brasilia, Brasilia, Brazil}\\
{\small $^{2}$Department of Economics, Federal University of Pelotas, Pelotas, Brazil}\\
{\small $^{3}$Facultad de Matem{\'a}ticas, Pontificia Universidad Cat{\'o}lica de Chile, Santiago, Chile}\\
}
\begin{document}
	\maketitle 	
\begin{abstract}
Power transformations of beta random variables produce unbiased estimating equations from transformed-model likelihood scores. This construction clarifies the connection between moment-type and likelihood-based estimating equations, recovers recent closed-form estimators for the beta shape parameters, and yields a new family indexed by a scalar transformation parameter $r$. For each fixed admissible $r$, we establish strong consistency and joint asymptotic normality. We also show that likelihood-guided selection over a prespecified finite grid preserves root-$n$ consistency and use one Fisher-scoring update to obtain an asymptotically efficient estimator. A Monte Carlo study with 1000 replications compares numerical maximum likelihood, two existing closed-form procedures, the proposed likelihood-selected closed-form estimator, and its one-step refinement across nine parameter configurations and four sample sizes. All procedures are evaluated on the same simulated samples within each replication. The selected closed-form estimator closely tracks maximum likelihood, while the one-step refinement is nearly indistinguishable from it for moderate sample sizes. Finally, the methods are illustrated descriptively using the proportions of municipal area devoted to crops and pasture in the 15 municipalities of Roraima, Brazil.

\end{abstract}



\section{Introduction}
\noindent
Closed-form point estimators are attractive wherever computational speed, robustness to initialization, or guaranteed convergence are at stake. By contrast, maximum likelihood (ML) for multi-parameter families often requires iterative optimization and may be sensitive to starting values, flat likelihood regions, or boundary estimates. This is particularly salient within exponential families, where, despite their elegant structure, ML estimators rarely admit explicit formulas once we move beyond the one-parameter case.

A classical way to bypass iterative ML is to use moment-type estimators. For exponential families, \citet{Davidson1974} linked moment-type estimation to likelihood equations in canonical parametrizations. A particularly influential development was the likelihood-equation construction of \citet{YCh2016} for the two-parameter gamma model. Subsequent work derived finite-sample bias corrections for those estimators \citep{LouzadaEtAl2019}, analytical maximum a posteriori alternatives \citep{LouzadaRamos2018}, and score-adjusted formulations that also cover the beta distribution \citep{Tamae2020}.

The search for explicit estimators is not confined to the gamma model. For Nakagami models, the progression from likelihood-based and generalized-moment procedures \citep{Cheng-Beaulieu2001,Cheng-Beaulieu2002} to closed-form MAP, bias-corrected, and MLE-like estimators is documented in \citet{RLR2016}, \citet{RamosEtAl2020}, and \citet{Zhao2021}. Related constructions have been developed for the weighted Lindley family \citep{GhitanyEtAl2017,Kim2021}, for transformation-based families containing gamma-related and weighted-Lindley-related models \citep{Kim2022}, and for matrix-variate or multivariate gamma models \citep{Alfelt2020,ZhaoJangKim2022,OikonomidisTrevezas2025}. Together, these studies show that suitable transformations, moments, or auxiliary likelihood equations can replace iterative optimization without giving up consistency and asymptotic normality.

At a broader methodological level, \citet{RamosEtAl2021} introduced generalized closed-form maximum likelihood equations and studied their invariance, consistency, and asymptotic normality, with illustrations for gamma, Nakagami, beta, and bivariate gamma distributions. For exponential-family models, \citet{Vila2024} derived closed-form estimators directly from likelihood equations, while \citet{Vila2024b} and \citet{VilaSaulo2024} extended likelihood-based and moment-type constructions to a weighted exponential family. These papers are especially close to the present work in their use of auxiliary models and transformed statistics. The construction developed here is distinct in that it starts from power-transformed beta variables, uses scores with respect to the transformation index to produce unbiased estimating equations, and then selects among a finite family of explicit solutions by the beta likelihood.

For the beta distribution itself, classical estimation includes moment procedures, numerical ML, interval estimation, and bias-corrected ML \citep{LauLau1991,CordeiroEtAl1997,GuptaNadarajah2004}. The recent closed-form literature follows several complementary routes. \citet{Tamae2020} used score adjustment; \citet{NawaNadarajah2023} proposed additional explicit estimators; \citet{Papadatos2024} used Stein-type covariance identities and U-statistics; \citet{NikWeiss2024} developed generalized moment estimators based on Stein identities; and \citet{Chen-Xiao2025} combined mixed moments with score equations from generalized beta models. This progression provides the immediate context for the transformed-score family proposed below.

{In this paper, we use power transformations $Y_j=[g_j^{-1}(X)]^{1/p}$, $j=1,2$, where $X \sim \mathrm{Beta}(\boldsymbol{\theta})$, $\boldsymbol{\theta}=(\alpha,\beta)^\top$, and $g_j$ is monotone and twice differentiable. The likelihood score in the auxiliary parameter $p$ induces unbiased estimating equations for $(\alpha,\beta)$. Because the two equations may arise from two different transformed models, they are not the score of a single joint likelihood; throughout the paper they are therefore treated as transformed-score estimating equations.}

For the induced model, we show that the score components can be written as empirical minus population functionals, leading to estimating equations with the same fixed points as the likelihood equations. Then, we can
\begin{enumerate}[(i)]
\item construct closed-form estimators by choosing convenient functions $g_j$ that linearize the score components in the unknown parameters;
\item recover known closed-form estimators as special cases (for instance, \citet{Chen-Xiao2025} and \citet{Tamae2020} for the beta model);
\item {generate a new family indexed by the transformation parameter $r$ in the beta illustration;}
\item {establish strong consistency and a joint central limit theorem for each fixed admissible transformation and root-$n$ consistency after selection over a finite grid.}
\end{enumerate}

We take the beta distribution as a detailed example. Using a general score identity with specific choices of $g$, we directly retrieve the closed-form estimators of \citet{Chen-Xiao2025} and \citet{Tamae2020}. {We then use $g_1(x)=x$, $0<x<1$, and $g_2(x)=(1-x^{1/s})^{1/r}$, $0<x<1$, $r,s>0$. Although the auxiliary transformation contains both $r$ and $s$, the parameter $s$ cancels from the resulting estimating equation. Thus the proposed closed-form family is indexed only by $r$. We compute the estimator over a finite grid of $r$ values and retain the admissible candidate with the largest beta log-likelihood. This is likelihood-guided selection among estimating-equation solutions, rather than profile likelihood in the usual parametric sense.}

The rest of this paper proceeds as follows. In Section \ref{sec:prel}, we present some preliminary definitions which will be used throughout the paper. In Section \ref{sec:closed},  we show that moment-type estimators of the population parameters can be derived from the likelihood equations for distributions within the exponential family. 
In this section, we also show that our approach has as special cases the closed-form estimators of \citet{Chen-Xiao2025} and \citet{Tamae2020}. In Section \ref{New estimators}, based on the results of Section \ref{sec:closed}, we propose new closed-form estimators for the parameters indexing the beta distribution. In Section \ref{Asymptotic behavior of estimators}, we derive a general class of closed-form estimators for the Beta distribution, and we establish their asymptotic behavior.
In Section \ref{sec:asymptotically_efficient}, we propose new asymptotically efficient estimators, based on those discussed in Section \ref{New estimators}.
In Section~\ref{sec:sim}, we present a Monte Carlo simulation study to evaluate the finite-sample performance of the proposed estimators for the beta model. In Section~\ref{Sec:application}, we develop an empirical application to municipal farming shares in Roraima (Brazil). Finally, in Section \ref{sec:concluding}, we provide some concluding remarks, and the Appendix presents some auxiliary results.

	\section{The Beta distribution}\label{sec:prel}

	\noindent

		The Beta distribution is an absolutely continuous probability distribution supported on the unit interval $(0,1)$, widely used in the modeling of proportions and probabilities.
		
		A random variable $X$ is said to follow a Beta distribution with shape parameters $\alpha>0$ and $\beta>0$, denoted by $X \sim \mathrm{Beta}(\boldsymbol{\theta})$, $\boldsymbol{\theta}=(\alpha,\beta)^\top$, if its probability density function is given by
		\begin{equation*}
			f(x;\boldsymbol{\theta})
			=
			\frac{1}{B(\alpha,\beta)} \,  x^{\alpha-1}(1-x)^{\beta-1}, 
			\quad 0 < x < 1,
		\end{equation*}
		where $B(\alpha,\beta)$ is the Beta function, defined as
		\begin{equation*}
			B(\alpha,\beta)
			=
			\int_0^1 t^{\alpha-1}(1-t)^{\beta-1}\,{\rm d}t.
		\end{equation*}
		
		An equivalent representation of the Beta function in terms of the Gamma function is
		\begin{equation*}
			B(\alpha,\beta)
			=
			\frac{\Gamma(\alpha)\Gamma(\beta)}{\Gamma(\alpha+\beta)}.
		\end{equation*}
		
		The first two moments of the Beta distribution are well known. The mean and variance are given by
		\begin{equation*}
			\mathbb{E}[X] = \frac{\alpha}{\alpha+\beta}, 
			\quad
			\mathrm{Var}(X)
			=
			\frac{\alpha\beta}{(\alpha+\beta)^2(\alpha+\beta+1)}.
		\end{equation*}
		
		The flexibility of the Beta distribution arises from the shape parameters $\alpha$ and $\beta$, which control symmetry and skewness. In particular, when $\alpha=\beta=1$, the distribution reduces to the uniform distribution on $(0,1)$, while different choices of parameters allow for a wide range of shapes, including symmetric, skewed, and U-shaped densities.

It is well known that the maximum likelihood (ML) estimators for the Beta distribution do not admit closed-form expressions; however, closed-form estimators are available via the method of moments \citep{LauLau1991,GuptaNadarajah2004}. In the next section, we extend the Beta distribution to a broader class through power transformations, with the aim of deriving new closed-form estimators for the population parameters. 

\smallskip 


Let  $X \sim \mathrm{Beta}(\boldsymbol{\theta})$.
By using the power transformations 
\begin{align}\label{def-Y-g}
Y_j = \bigl[g_j^{-1}(X)\bigr]^{1/p}, \quad j=1,2, \ p > 0,
\end{align} 
where $g_j:D_j\subset (0,\infty)\to (0,1)$ is a twice-differentiable monotonic function with inverse $g_j^{-1}$, it is simple to see that the PDF of $Y$ is written as
%
\begin{align}\label{A generalized exponential family}
	f_{Y_j}(y;\boldsymbol{\theta},p)
	=
\frac{p}{B(\alpha,\beta)} \,  
g_j^{\alpha-1}(y^p) y^{p-1} [1-g_j(y^p)]^{\beta-1}
\vert g_j'(y^p)\vert.
\end{align}
	In the above, $g'_j(x)$ denotes the derivative of $g_j(x)$ with respect to $x$.

\section{{Transformed likelihood scores and estimating equations for the power transformation model \eqref{def-Y-g}}}\label{sec:closed}

For a fixed transformation $g_j$, the equation below is the score equation in the
auxiliary power parameter $p$ of the corresponding transformed model. When two
different transformations are combined to estimate $(\alpha,\beta)$, the resulting
pair is an unbiased estimating system; it is not the gradient of one joint
log-likelihood.

	\noindent
	
Let $\{Y_{ji} : i = 1,\ldots, n\}$ be a univariate random sample of size $n$ from $Y_j$ having PDF \eqref{A generalized exponential family}, for $j=1,2$.
The corresponding (random) log-likelihood function for $(\boldsymbol{\theta},p)^\top$ is given by
\begin{align*}
	l_j(\boldsymbol{\theta},p)
	&=
	n\log(p)-n\log(B(\alpha,\beta))
	+
	(\alpha-1)
	\sum_{i=1}^{n}
	\log(g_j(Y_{ji}^p))
	\\[0,2cm]
	&
	+
	(p-1)
	\sum_{i=1}^{n}
	\log(Y_{ji})
	+
	(\beta-1) 
	\sum_{i=1}^{n}
	\log(1-g_j(Y_{ji}^p))
	+
	\sum_{i=1}^{n}
	\log(\vert g_j'(Y_{ji}^p)\vert).
\end{align*}

The (random) score vector is given by the following partial derivatives:
\begin{align*}
	&{\partial l_j(\boldsymbol{\theta},p)\over\partial\alpha}
	=
	-
n[\psi(\alpha)-\psi(\alpha+\beta)]
	+
	\sum_{i=1}^{n}
\log(g_j(Y_{ji}^p)),
	\\[0,2cm]
&{\partial l_j(\boldsymbol{\theta},p)\over\partial\beta}
=
-
n[\psi(\beta)-\psi(\alpha+\beta)]
+
\sum_{i=1}^{n}
\log(1-g_j(Y_{ji}^p)),
	\\[0,2cm]
	&{\partial l_j(\boldsymbol{\theta},p)\over\partial p}
	=
	{1\over p} \, 
	\Bigg[
	n
	+
(\alpha - 1)
\sum_{i=1}^{n}
\frac{g_j'(Y_{ji}^p)}{g_j(Y_{ji}^p)}
\, Y_{ji}^p \log(Y_{ji}^p)
			+
	\sum_{i=1}^{n}
	\log(Y_{ji}^p)
	\nonumber
		\\[0,2cm]
	&\hspace*{1.7cm}
-
(\beta - 1)
\sum_{i=1}^{n}
\frac{g_j'(Y_{ji}^p)}{1 - g_j(Y_{ji}^p)}
\, Y_{ji}^p \log(Y_{ji}^p)
	+
	\sum_{i=1}^{n}
	 \frac{g_j''(Y_{ji}^p)}{g_j'(Y_{ji}^p)}\, Y_{ji}^p \log(Y_{ji}^p)
	 \Bigg],
\end{align*}
where $\psi(x)=\Gamma'(x)/\Gamma(x)$ is the digamma function.
Since 
$$
g_j(Y_{ji}^p)=X_i \quad \text{and} \quad  Y_{ji}^p=g_j^{-1}(X_i), \quad i=1,\ldots,n,
$$ 
where $\{X_i : i = 1,\ldots,n\}$ is a univariate random sample of size $n$ from $X \sim \mathrm{Beta}(\boldsymbol{\theta})$, the above equations can be expressed in terms of $X_1,\ldots,X_n$ as follows:
\begin{align*}
	&{\partial l_j(\boldsymbol{\theta},p)\over\partial\alpha}
	=
	-
	n[\psi(\alpha)-\psi(\alpha+\beta)]
	+
	\sum_{i=1}^{n}
	\log(X_i),
	\\[0,2cm]
	&{\partial l_j(\boldsymbol{\theta},p)\over\partial\beta}
	=
	-
	n[\psi(\beta)-\psi(\alpha+\beta)]
	+
	\sum_{i=1}^{n}
	\log(1-X_i),
	\\[0,2cm]
	&{\partial l_j(\boldsymbol{\theta},p)\over\partial p}
	=
	{1\over p} \, 
	\Bigg[
	n
	+
	(\alpha - 1)
	\sum_{i=1}^{n}
	\frac{g_j'\bigl(g_j^{-1}(X_i)\bigr)}{X_i}
	\, g_j^{-1}(X_i) \log\bigl(g_j^{-1}(X_i)\bigr)
	+
	\sum_{i=1}^{n}
	\log\bigl(g_j^{-1}(X_i)\bigr)
	\nonumber
	\\[0,2cm]
	&\hspace*{1.7cm}
	-
	(\beta - 1)
	\sum_{i=1}^{n}
	\frac{g_j'\bigl(g_j^{-1}(X_i)\bigr)}{1 - X_i}
	\, g_j^{-1}(X_i) \log\bigl(g_j^{-1}(X_i)\bigr)
	\\[0,2cm]
	&\hspace*{1.7cm}
	+
	\sum_{i=1}^{n}
	\frac{g_j''\bigl(g_j^{-1}(X_i)\bigr)}{g_j'\bigl(g_j^{-1}(X_i)\bigr)}\, g_j^{-1}(X_i) \log\bigl(g_j^{-1}(X_i)\bigr)
	\Bigg].
\end{align*}

Hence,
\begin{align}
	{\partial l_j(\boldsymbol{\theta},p)\over \partial\alpha}
	=
	0
	\quad  \Longleftrightarrow \quad &
\psi(\alpha)-\psi(\alpha+\beta)
=
{1\over n}
\sum_{i=1}^{n}
\log(X_i),
	\label{f-eq}
	\\[0,2cm]
	{\partial l_j(\boldsymbol{\theta},p)\over \partial\beta}
=
0
\quad  \Longleftrightarrow \quad &
\psi(\beta)-\psi(\alpha+\beta)
=
{1\over n}
\sum_{i=1}^{n}
\log(1-X_i),
\label{f-eq-1}	
	\\[0,2cm]
	{\partial l_j(\boldsymbol{\theta},p)\over \partial p}
=
0
\quad  \Longleftrightarrow \quad &		
(\alpha - 1) \,
{1\over n}
\sum_{i=1}^{n}
\frac{g_j'\bigl(g_j^{-1}(X_i)\bigr)}{X_i}
\, g_j^{-1}(X_i) \log\bigl(g_j^{-1}(X_i)\bigr)
\nonumber
\\[0,2cm]
&
-
(\beta - 1) \, 
{1\over n}
\sum_{i=1}^{n}
\frac{g_j'\bigl(g_j^{-1}(X_i)\bigr)}{1 - X_i}
\, g_j^{-1}(X_i) \log\bigl(g_j^{-1}(X_i)\bigr)
\nonumber
\\[0,2cm]
&
=
-
1
-
{1\over n}
\sum_{i=1}^{n}
\log\bigl(g_j^{-1}(X_i)\bigr)
-
{1\over n}
\sum_{i=1}^{n}
\frac{g_j''\bigl(g_j^{-1}(X_i)\bigr)}{g_j'\bigl(g_j^{-1}(X_i)\bigr)}\, g_j^{-1}(X_i) \log\bigl(g_j^{-1}(X_i)\bigr).
\label{s-eq}
\end{align}

It is worth noting that the equations in \eqref{f-eq}-\eqref{f-eq-1} do not yield closed-form solutions for $\alpha$ and $\beta$.

To obtain explicit expressions for $\alpha$ and $\beta$, 
in Examples \ref{ex:closedbeta} and \ref{ex:Tamae2020}, and in Section \ref{New estimators}, we use Equation \eqref{s-eq} with appropriate choices of twice-differentiable monotonic functions $g_j:D_j\subset (0,\infty)\to (0,1)$, $j=1,2$.

In the following examples, we show how to obtain the estimators proposed in \cite{Chen-Xiao2025} and \cite{Tamae2020}.
    \begin{example}[Obtaining the estimators proposed by \cite{Chen-Xiao2025}]\label{ex:closedbeta}
        	Let $g_1 : D_1 \equiv (0,1) \to (0,1)$ be the identity function, that is, 
	$$g_1(x)=x, \quad 0<x<1.$$ 
	Then
	\begin{align*}
		g_1^{-1}(x)=x, 
		\quad
		g_1'\bigl(g_1^{-1}(x)\bigr)=1,
		\quad
		\frac{g_1''\bigl(g_1^{-1}(x)\bigr)}{g_1'\bigl(g_1^{-1}(x)\bigr)}=0.
	\end{align*}
	Substituting these identities into \eqref{s-eq}, we obtain
	\begin{align*}
		(\alpha - 1)\,
		\frac{1}{n}\sum_{i=1}^{n}\log(X_i)
		-
		(\beta - 1)\,
		\frac{1}{n}\sum_{i=1}^{n}
		\frac{X_i}{1 - X_i}\log(X_i)
		=
		-1
		-
		\frac{1}{n}\sum_{i=1}^{n}\log(X_i).
	\end{align*}
	Equivalently,
	\begin{align}
		\label{eqq1}
		\alpha 
		\left[
		\frac{1}{n}\sum_{i=1}^{n}\log(X_i)
		\right]
		-
		\beta
		\left[
		\frac{1}{n}\sum_{i=1}^{n}
		\frac{X_i}{1-X_i}\log(X_i)
		\right]
		=
		-1
		-
		\frac{1}{n}\sum_{i=1}^{n}
		\frac{X_i}{1-X_i}\log(X_i).
	\end{align}
	
	\medskip\medskip\medskip
	
	Next, let $g_2 : D_2 \equiv (0,1) \to (0,1)$ be defined by $$g_2(x)=1-x, \quad 0<x<1.$$ 
	Then
	\begin{align*}
		g_2^{-1}(x)=1-x,
		\quad
		g_2'\bigl(g_2^{-1}(x)\bigr)=-1,
		\quad
		\frac{g_2''\bigl(g_2^{-1}(x)\bigr)}{g_2'\bigl(g_2^{-1}(x)\bigr)}=0.
	\end{align*}
	Substituting into \eqref{s-eq} yields
	\begin{align*}
		-(\alpha - 1)\,
		\frac{1}{n}\sum_{i=1}^{n}
		\frac{1-X_i}{X_i}\log(1-X_i)
		+
		(\beta - 1)\,
		\frac{1}{n}\sum_{i=1}^{n}\log(1-X_i)
		=
		-1
		-
		\frac{1}{n}\sum_{i=1}^{n}\log(1-X_i).
	\end{align*}
	Equivalently,
	\begin{align}
		\label{eqq2}
		-
		\alpha
		\left[
		\frac{1}{n}\sum_{i=1}^{n}
		\frac{1-X_i}{X_i}\log(1-X_i)
		\right]
		+
		\beta
		\left[
		\frac{1}{n}\sum_{i=1}^{n}\log(1-X_i)
		\right]
		=
		-1
		-
		\frac{1}{n}\sum_{i=1}^{n}
		\frac{1-X_i}{X_i}\log(1-X_i).
	\end{align}
	
	From \eqref{eqq1} and \eqref{eqq2}, we obtain the following system of equations:
	\begin{align*}
		\begin{cases}
			\displaystyle
			\alpha 
			\left[
			\frac{1}{n}\sum_{i=1}^{n}\log(X_i)
			\right]
			-
			\beta 
			\left[
			\frac{1}{n}\sum_{i=1}^{n}
			\frac{X_i}{1-X_i}\log(X_i)
			\right]
			&=\ \ \displaystyle
			-1
			-
			\frac{1}{n}\sum_{i=1}^{n}
			\frac{X_i}{1-X_i}\log(X_i),
			\\[0.7cm]
			\displaystyle
			-
			\alpha 
			\left[
			\frac{1}{n}\sum_{i=1}^{n}
			\frac{1-X_i}{X_i}\log(1-X_i)
			\right]
			+
			\beta 
			\left[
			\frac{1}{n}\sum_{i=1}^{n}\log(1-X_i)
			\right]
			&= \ \ \displaystyle
			-1
			-
			\frac{1}{n}\sum_{i=1}^{n}
			\frac{1-X_i}{X_i}\log(1-X_i).
		\end{cases}
	\end{align*}

For a compact expression of the same solution, define
\begin{align*}
L_X&=\frac1n\sum_{i=1}^n\log X_i,
&L_Y&=\frac1n\sum_{i=1}^n\log(1-X_i),\\
M_X&=\frac1n\sum_{i=1}^nX_i\log\!\left(\frac{X_i}{1-X_i}\right),
&M_Y&=\frac1n\sum_{i=1}^n(1-X_i)\log\!\left(\frac{1-X_i}{X_i}\right).
\end{align*}
Solving the system gives
\begin{align*}
\widehat\alpha
&=\frac{(1+M_X)L_Y+(1+M_Y)M_X}{M_XM_Y-L_XL_Y},\\
\widehat\beta
&=\frac{(1+M_Y)L_X+(1+M_X)M_Y}{M_XM_Y-L_XL_Y}.
\end{align*}
These are the closed-form estimators proposed by \citet{Chen-Xiao2025}.

    \end{example}

      \begin{example}[Obtaining the estimators proposed by \cite{Tamae2020}]\label{ex:Tamae2020}
        Let $g_1 : D_1 \equiv (0,\infty) \to (0,1)$ be defined by
\[
{g_1(x)=\frac{x}{x+1}, \quad x>0.}
\]
Then
\begin{align*}
	g_1^{-1}(x)=\frac{x}{1-x},
	\quad
	g_1'\bigl(g_1^{-1}(x)\bigr)=(1-x)^2,
	\quad
	\frac{g_1''\bigl(g_1^{-1}(x)\bigr)}{g_1'\bigl(g_1^{-1}(x)\bigr)}=-2(1-x).
\end{align*}
Substituting these identities into \eqref{s-eq}, we obtain
\begin{align*}
	(\alpha - 1)\,
	\frac{1}{n}\sum_{i=1}^{n}
	(1-X_i)\log\!\left(\frac{X_i}{1-X_i}\right)
	&-
	(\beta - 1)\,
	\frac{1}{n}\sum_{i=1}^{n}
	X_i\log\!\left(\frac{X_i}{1-X_i}\right)
	\\[0.2cm]
	&=
	-1
	-
	\frac{1}{n}\sum_{i=1}^{n}
	\log\!\left(\frac{X_i}{1-X_i}\right)
	+
	\frac{2}{n}\sum_{i=1}^{n}
	X_i\log\!\left(\frac{X_i}{1-X_i}\right).
\end{align*}
Equivalently,
\begin{align}
	\label{eqq3}
	\alpha
	\left[
	\frac{1}{n}\sum_{i=1}^{n}
	(1-X_i)\log\!\left(\frac{X_i}{1-X_i}\right)
	\right]
	-
	\beta
	\left[
	\frac{1}{n}\sum_{i=1}^{n}
	X_i\log\!\left(\frac{X_i}{1-X_i}\right)
	\right]
	=
	-1.
\end{align}

\medskip\medskip\medskip

Next, let $g_2 : D_2 \equiv (0,1) \to (0,1)$ be defined by
\[
g_2(x)=\frac{-\log(x)}{1-\log(x)}, \quad 0<x<1.
\]
Then
\begin{align*}
	g_2^{-1}(x)&=\exp\!\left(-\frac{x}{1-x}\right),\\
	g_2'\bigl(g_2^{-1}(x)\bigr)
	&=-(1-x)^2\exp\!\left(\frac{x}{1-x}\right),\\
	\frac{g_2''\bigl(g_2^{-1}(x)\bigr)}{g_2'\bigl(g_2^{-1}(x)\bigr)}
	&=
	-(2x-1)\exp\!\left(\frac{x}{1-x}\right).
\end{align*}
Substituting into \eqref{s-eq} yields
\begin{align*}
	(\alpha - 1)\,
	\frac{1}{n}\sum_{i=1}^{n}
	(1-X_i)
	-
	(\beta - 1)\,
	\frac{1}{n}\sum_{i=1}^{n}
	X_i
	=
	-1
	+
	\frac{1}{n}\sum_{i=1}^{n}
	\frac{X_i}{1-X_i}
	-
	\frac{1}{n}\sum_{i=1}^{n}
	(2X_i-1)\frac{X_i}{1-X_i}.
\end{align*}
Equivalently,
\begin{align}
	\label{eqq4}
	\alpha
	\left[
	\frac{1}{n}\sum_{i=1}^{n}(1-X_i)
	\right]
	-
	\beta
	\left[
	\frac{1}{n}\sum_{i=1}^{n}X_i
	\right]
	=
	0.
\end{align}

From \eqref{eqq3} and \eqref{eqq4}, we obtain the system
\begin{align*}
	\begin{cases}
		\displaystyle
		\alpha
		\left[
		\frac{1}{n}\sum_{i=1}^{n}
		(1-X_i)\log\!\left(\frac{X_i}{1-X_i}\right)
		\right]
		-
		\beta
		\left[
		\frac{1}{n}\sum_{i=1}^{n}
		X_i\log\!\left(\frac{X_i}{1-X_i}\right)
		\right]
		&= \ \ -1,
		\\[0.7cm]
		\displaystyle
		\alpha
		\left[
		\frac{1}{n}\sum_{i=1}^{n}(1-X_i)
		\right]
		-
		\beta
		\left[
		\frac{1}{n}\sum_{i=1}^{n}X_i
		\right]
		&= \ \ 0.
	\end{cases}
\end{align*}

Solving this system yields the estimators for $\alpha$ and $\beta$:
\begin{align*}
	\widehat{\alpha}
	=
	\frac{\displaystyle \frac{1}{n}\sum_{i=1}^{n}X_i}{
		\displaystyle
		\frac{1}{n}\sum_{i=1}^{n}
		X_i\log\!\left(\frac{X_i}{1-X_i}\right)
		-
		\left(\frac{1}{n}\sum_{i=1}^{n}X_i\right)
		\left[\frac{1}{n}\sum_{i=1}^{n}
		\log\!\left(\frac{X_i}{1-X_i}\right)\right]
	},
\end{align*}
and
\begin{align*}
	\widehat{\beta}
	=
	\frac{\displaystyle 1-\frac{1}{n}\sum_{i=1}^{n}X_i}{
		\displaystyle
		\frac{1}{n}\sum_{i=1}^{n}
		X_i\log\!\left(\frac{X_i}{1-X_i}\right)
		-
		\left(\frac{1}{n}\sum_{i=1}^{n}X_i\right)
		\left[\frac{1}{n}\sum_{i=1}^{n}
		\log\!\left(\frac{X_i}{1-X_i}\right)\right]
	}.
\end{align*}

These estimators are identical to those introduced in \cite{Tamae2020}.

    \end{example}

\section{{A new closed-form family for $\alpha$ and $\beta$}}\label{New estimators}

	\noindent

In this section, we introduce a new pair of functions, $g_1$ and $g_2$, which enable the derivation of a new class of closed-form estimators for the parameters of the Beta distribution.

As in the derivation of \eqref{eqq1}, taking $g_1:D_1\equiv(0,1)\to(0,1)$ as the identity function and substituting into \eqref{s-eq} yields
\begin{align}
	\label{eqq1-new}
	\alpha 
	\left[
	\frac{1}{n}\sum_{i=1}^{n}\log(X_i)
	\right]
	-
	\beta
	\left[
	\frac{1}{n}\sum_{i=1}^{n}
	\frac{X_i}{1-X_i}\log(X_i)
	\right]
	=
	-1
	-
	\frac{1}{n}\sum_{i=1}^{n}
	\frac{X_i}{1-X_i}\log(X_i).
\end{align}

\medskip\medskip\medskip

Next, consider the transformation $g_2:D_2\equiv(0,1)\to(0,1)$ defined by
\[
g_2(x)=(1-x^{1/s})^{1/r}, \quad r,s>0.
\]
Then
\begin{align*}
	g_2^{-1}(x)&=(1-x^r)^s,\\[0,2cm]
	g_2'\bigl(g_2^{-1}(x)\bigr)&=-\frac{(1-x^r)^{1-s}x^{1-r}}{rs},\\[0,2cm]
	\frac{g_2''\bigl(g_2^{-1}(x)\bigr)}{g_2'\bigl(g_2^{-1}(x)\bigr)}
	&=
	\frac{(1-x^r)^{-s}x^{-r}\,[\,r-1+(1-rs)x^r\,]}{rs}.
\end{align*}
Substituting these expressions into \eqref{s-eq} leads to
\begin{align}
	\label{eq12}
	&-
	\alpha
	\left[
	\frac{1}{n}\sum_{i=1}^{n}
	\frac{1-X_i^r}{X_i^{r}} \,
	\log(1-X_i^r)
	\right]
	+
	\beta
	\left[
	\frac{1}{n}\sum_{i=1}^{n}
	\frac{1-X_i^r}{(1-X_i)X_i^{r-1}} \,
	\log(1-X_i^r)
	\right]
	\nonumber
	\\[0.2cm]
	&=
	-r
	-
	\frac{1}{n}
	\sum_{i=1}^{n}
	{\frac{r-1+(1-rs)X_i^r}{X_i^r}}
	\,\log(1-X_i^r)
	\nonumber
	\\[0.2cm]
	&\quad
	-
	\frac{1}{n}
	\sum_{i=1}^{n}
	\frac{1-X_i^r}{X_i^{r}}
	\,\log(1-X_i^r)
	+
	\frac{1}{n}
	\sum_{i=1}^{n}
	\frac{1-X_i^r}{(1-X_i)X_i^{r-1}}
	\,\log(1-X_i^r)
	\nonumber
	\\[0.2cm]
	&\quad
	-
	rs
	\left[
	\frac{1}{n}\sum_{i=1}^{n}\log(1-X_i^r)
	\right].
\end{align}

Combining \eqref{eqq1-new} and \eqref{eq12}, we obtain the system
\begin{align*}
	\begin{cases}
		\displaystyle
		\alpha 
		\left[
		\frac{1}{n}\sum_{i=1}^{n}\log(X_i)
		\right]
		-
		\beta
		\left[
		\frac{1}{n}\sum_{i=1}^{n}
		\frac{X_i}{1-X_i}\log(X_i)
		\right]
		=
		\displaystyle
		-1
		-
		\frac{1}{n}\sum_{i=1}^{n}
		\frac{X_i}{1-X_i}\log(X_i),
		\\[0.9cm]
		\displaystyle
		-\alpha
		\left[
		\frac{1}{n}\sum_{i=1}^{n}
		\frac{1-X_i^r}{X_i^{r}} \,
		\log(1-X_i^r)
		\right]
		+
		\beta
		\left[
		\frac{1}{n}\sum_{i=1}^{n}
		\frac{1-X_i^r}{(1-X_i)X_i^{r-1}} \,
		\log(1-X_i^r)
		\right]
		\\[0.5cm]
		\displaystyle
		\qquad =
		-r
		-
		\frac{1}{n}
		\sum_{i=1}^{n}
		{\frac{r-1+(1-rs)X_i^r}{X_i^r}}
		\,\log(1-X_i^r)
		\\[0.5cm]
		\displaystyle
		\qquad\quad
		-
		\frac{1}{n}
		\sum_{i=1}^{n}
		\frac{1-X_i^r}{X_i^{r}}
		\,\log(1-X_i^r)
		+
		\frac{1}{n}
		\sum_{i=1}^{n}
		\frac{1-X_i^r}{(1-X_i)X_i^{r-1}}
		\,\log(1-X_i^r)
		\\[0.5cm]
		\displaystyle
		\qquad\quad
		-
		rs
		\left[
		\frac{1}{n}\sum_{i=1}^{n}\log(1-X_i^r)
		\right].
	\end{cases}
\end{align*}

Solving this system yields the following closed-form estimators for $\alpha$ and $\beta$:
\begin{align}
	\label{closed_alpha}
	\widehat{\alpha}_r
	=
	{\frac{(\widehat{\beta}_r-1)B - 1}{A}},
\end{align}
and
\begin{align}
	\label{closed_beta}
	\widehat{\beta}_r
	=
	{\frac{-r-(r-1)G-F-C+D-\dfrac{(B+1)C}{A}}
	{D-\dfrac{BC}{A}}},
\end{align}
where
\begin{align*}
	A&=\frac{1}{n}\sum_{i=1}^{n}\log(X_i),\\[0.2cm]
	B&=\frac{1}{n}\sum_{i=1}^{n}\frac{X_i}{1-X_i}\log(X_i),\\[0.2cm]
	C&=\frac{1}{n}\sum_{i=1}^{n}\frac{1-X_i^r}{X_i^r}\log(1-X_i^r),\\[0.2cm]
	D&=\frac{1}{n}\sum_{i=1}^{n}\frac{1-X_i^r}{(1-X_i)X_i^{r-1}}\log(1-X_i^r),\\[0.2cm]
	{E^\star}&{=\frac{1}{n}\sum_{i=1}^{n}
	\frac{r-1+(1-rs)X_i^r}{X_i^r}
	\log(1-X_i^r)=(r-1)G+(1-rs)F,}\\[0.2cm]
	F&=\frac{1}{n}\sum_{i=1}^{n}\log(1-X_i^r).
\\[-0.1cm]
	{G}&{=\frac{1}{n}\sum_{i=1}^{n}\frac{\log(1-X_i^r)}{X_i^r}.}
\end{align*}

The identity $E^\star=(r-1)G+(1-rs)F$ shows that
$-E^\star-rsF=-(r-1)G-F$. Hence $s$ cancels from the estimating
equation and the resulting family is indexed by $r$ alone. This cancellation is
the reason for writing the estimators as $(\widehat\alpha_r,\widehat\beta_r)$.

\begin{remark}\label{remark:profile}
For a finite grid $\mathcal R\subset(0,\infty)$, let $\mathcal A_n$ contain the
values of $r$ for which \eqref{closed_alpha}-\eqref{closed_beta} produce finite,
strictly positive estimates. 
    We define
\begin{equation}\label{proposed_estimator}
 \widehat r\in\arg\max_{r\in\mathcal A_n}\;
 \ell_n\!\left(\widehat\alpha_r,\widehat\beta_r\right)~~\text{and}~~(\widehat\alpha,\widehat\beta):=
(\widehat\alpha_{\widehat r},\widehat\beta_{\widehat r}).
\end{equation}
We use $\mathcal R=\{0.1,0.2,\ldots,2.5\}$ and break likelihood ties by choosing the smallest $r$. Because $r$ is not a parameter of the beta distribution, this is likelihood-guided selection among a finite collection of estimating-equation solutions, not profile likelihood in the conventional sense.
\end{remark}

\begin{remark}
{For the particular case $r=1$, we have $D=F$, leading to the simplified expressions for the estimators of}
$\alpha$ and $\beta$:
\begin{align*}
	\widehat{\alpha}_1
	=
	{{(\widehat{\beta}_1-1) B-1 \over A}},
\end{align*}
and 
\begin{align*}
	\widehat{\beta}_1
	=
	{{\displaystyle -1-C-{(B+1)C\over A}\over\displaystyle D-{BC\over A}}},
\end{align*}
respectively, 
which coincide with the estimators (of 
$\alpha$ and $\beta$) presented in Example \ref{ex:closedbeta}. 
\end{remark}

\section{A general class of closed-form estimators and its asymptotic behavior}
\label{Asymptotic behavior of estimators}\label{sect-8}

In this section, we derive a general class of closed-form estimators for the Beta distribution (Eqs. \eqref{eq:estimator_general1} and \eqref{eq:estimator_general2}), and we establish their strong consistency and a central limit theorem (Theorem \ref{clt}).
As a particular case, we show in Theorem \ref{clt-1} that conditions of Theorem \ref{clt} are satisfied for the new estimator $(\widehat{\alpha}, \widehat{\beta})^\top$ proposed in Section \ref{New estimators}.

	Let $\{X_i : i = 1,\ldots,n\}$ be a random sample of $X \sim \mathrm{Beta}(\boldsymbol{\theta})$.
	For notational convenience, define
\begin{align}\label{defab1}
	\overline{X}_{kj}
	\equiv
	\frac{1}{n}\sum_{i=1}^{n} 
	h_{kj}(X_i),
	\quad k=1,2,3,4,5,\; j=1,2,
\end{align}
where
\begin{align}\label{defab2}
	\begin{aligned}
		h_{1j}(x)&\equiv \frac{1}{x}\,
		g_j'(g_j^{-1}(x))\, g_j^{-1}(x)\log\bigl(g_j^{-1}(x)\bigr),\\[0.3cm]
		h_{2j}(x)&\equiv \frac{1}{1-x}\,
		g_j'(g_j^{-1}(x))\, g_j^{-1}(x)\log\bigl(g_j^{-1}(x)\bigr),\\[0.3cm]
		h_{3j}(x)&\equiv 
		\frac{g_j''(g_j^{-1}(x))}{g_j'(g_j^{-1}(x))}\,
		g_j^{-1}(x)\log\bigl(g_j^{-1}(x)\bigr),\\[0.3cm]
		h_{4j}(x)&\equiv 
		\frac{(2x-1)\, g_j'(g_j^{-1}(x))}{x(1-x)}\,
		g_j^{-1}(x)\log\bigl(g_j^{-1}(x)\bigr),\\[0.3cm]
		h_{5j}(x)&\equiv \log\bigl(g_j^{-1}(x)\bigr),
	\end{aligned}
\end{align}
and $g_j:D_j\subset(0,\infty)\to(0,1)$, $j=1,2$, are monotone and at least twice continuously differentiable functions, as defined in \eqref{def-Y-g}.

With this notation, Equation \eqref{s-eq} can be rewritten as
\begin{align}\label{defab}
	\alpha \overline{X}_{1j}
	-
	\beta \overline{X}_{2j}
	=
	-1
	-
	\overline{X}_{3j}
	-
	\overline{X}_{4j}
	-
	\overline{X}_{5j},
	\quad j=1,2.
\end{align}

From system \eqref{defab}, the following estimators for $\alpha$ and $\beta$ are obtained:
\begin{align}\label{eq:estimator_general1}
	\widehat{\alpha}
	=
	\dfrac
	{ \displaystyle
		{1+\overline{X}_{31}+\overline{X}_{41}+\overline{X}_{51}\over \overline{X}_{21}}\, \overline{X}_{22}
		-
		1-\overline{X}_{32}-\overline{X}_{42}-\overline{X}_{52}
	}
	{ \displaystyle
		\overline{X}_{12}-{\overline{X}_{11}\overline{X}_{22}\over \overline{X}_{21}}
	},
\end{align}
and
\begin{align}\label{eq:estimator_general2}
	\widehat{\beta}
	=
	\dfrac{\widehat{\alpha}\, \overline{X}_{11}+1+\overline{X}_{31}+\overline{X}_{41}+\overline{X}_{51}}{\overline{X}_{21}},
\end{align}
respectively.

Furthermore, consider the notation for the $10\times1$ vectors
\begin{align}\label{defabnew}
	\overline{\boldsymbol{X}}
\equiv
	\begin{pmatrix}
\overline{X}_{1j}&\\
\overline{X}_{2j}&\\
\overline{X}_{3j}&: \ j=1,2\\
\overline{X}_{4j}&\\
\overline{X}_{5j}&
	\end{pmatrix},
	\quad 
\boldsymbol{X}
\equiv
	\begin{pmatrix}
h_{1j}(X)&\\
h_{2j}(X)&\\
h_{3j}(X)&: \ j=1,2\\
h_{4j}(X)&\\
h_{5j}(X)&
	\end{pmatrix},
\end{align}
with $\overline{X}_{kj}$ and $h_{kj}$ defined in \eqref{defab1} and \eqref{defab2}, respectively.
By the strong law of large numbers, we have
\begin{align*}
	\overline{\boldsymbol{X}}\stackrel{\rm a.s.}{\longrightarrow}
	\mathbb{E}[\boldsymbol{X}],
\end{align*}
where ``$\stackrel{\rm a.s.}{\longrightarrow}$'' denotes almost sure convergence.

Note that
\begin{align}\label{id-1-0}
\widehat{\alpha}=\xi_1(\overline{\boldsymbol{X}})
	\quad \text{and} \quad
\widehat{\beta}=\xi_2(\overline{\boldsymbol{X}}),
\end{align}
with
\begin{align}\label{def-g1}
	\xi_1(\boldsymbol{x})
	\equiv
	\dfrac
	{\displaystyle
		{1+x_{31}+x_{41}+x_{51}\over x_{21}}\, x_{22}
		-
		1-x_{32}-x_{42}-x_{52}
	}
	{\displaystyle
		x_{12}-{x_{11}x_{22}\over x_{21}}
	},
\end{align}
\begin{align}\label{def-g2}
	\xi_2(\boldsymbol{x})
	\equiv
	\dfrac{
		\xi_1(\boldsymbol{x})
		x_{11}+1+x_{31}+x_{41}+x_{51}}{x_{21}},
\end{align}
and
\begin{align*}
	\boldsymbol{x}
	\equiv
	\begin{pmatrix}
		x_{1j} & \\
		x_{2j} & \\
		x_{3j} & : \ j=1,2  \\
		x_{4j} & \\
		x_{5j}
	\end{pmatrix}.
\end{align*}

Applying the continuous mapping Theorem \citep{Billingsley1969} yields
\begin{align}\label{id-1}
	\widehat{\alpha}\stackrel{\eqref{id-1-0}}{=}\xi_1(\overline{\boldsymbol{X}})
	\stackrel{\rm a.s.}{\longrightarrow}
	\xi_1(\mathbb{E}[\boldsymbol{X}])
	\quad \text{and} \quad
	\widehat{\beta}\stackrel{\eqref{id-1-0}}{=}\xi_2(\overline{\boldsymbol{X}})
	\stackrel{\rm a.s.}{\longrightarrow}
	\xi_2(\mathbb{E}[\boldsymbol{X}]),
\end{align}

Moreover, under the moment conditions
$\mathbb{E}[h_{kj}^2(X)]<\infty$ for all $k=1,\ldots,5$ and $j=1,2$, it follows from the multivariate central limit theorem that
\begin{align*}
	\sqrt{n}\big\{\overline{\boldsymbol{X}}-\mathbb{E}[\boldsymbol{X}]\big\}\stackrel{\mathscr D}{\longrightarrow} N_{10}(\bm 0, \bm\Sigma),
\end{align*}
where $\bm\Sigma$ denotes the covariance matrix of $\boldsymbol{X}$ and ``$\stackrel{\mathscr D}{\longrightarrow}$'' means convergence in distribution.
Consequently, if the mapping $\boldsymbol{\xi}=(\xi_1,\xi_2)^\top$ is continuously differentiable in a neighborhood of $\mathbb{E}[\boldsymbol X]$ and the Jacobian matrix
\[
\boldsymbol{A}
=
\left.
\frac{\partial (\xi_1,\xi_2)}{\partial \boldsymbol x} \,
\right|_{\boldsymbol x=\mathbb{E}[\boldsymbol X]} 
\]
has full row rank, then the delta method implies
\begin{align}\label{id-2}
	\sqrt{n}
	\left\{
	\begin{pmatrix}
		\widehat{\alpha}
		\\[0,2cm]
		\widehat{\beta}
	\end{pmatrix}
	-
	\begin{pmatrix}
		\xi_1(\mathbb{E}[\bm X])
		\\[0,2cm]
		\xi_2(\mathbb{E}[\bm X])
	\end{pmatrix}
	\right\}
	\stackrel{\eqref{id-1-0}}{=}
	\sqrt{n}
	\left\{
	\begin{pmatrix}
		\xi_1(\overline{\boldsymbol{X}})
		\\[0,2cm]
		\xi_2(\overline{\boldsymbol{X}})
	\end{pmatrix}
	-
	\begin{pmatrix}
		\xi_1(\mathbb{E}[\bm X])
		\\[0,2cm]
		\xi_2(\mathbb{E}[\bm X])
	\end{pmatrix}
	\right\}
	\stackrel{\mathscr D}{\longrightarrow}
	N_2(\bm 0, \bm A\bm\Sigma \bm A^\top).
\end{align}

Under mild conditions, the following result establishes the strong consistency and a central limit theorem-type behavior for the estimators of $\alpha$ and $\beta$.

\begin{theorem}\label{clt}
	Let $\{X_i : i = 1,\ldots,n\}$ be a random sample of $X \sim \mathrm{Beta}(\boldsymbol{\theta})$ and define
	$
	\overline{\boldsymbol X}
	$
	as in \eqref{defabnew}. Let $(\widehat{\alpha},\widehat{\beta})^\top
	=
	\boldsymbol{\xi}(\overline{\boldsymbol X})$, where $\boldsymbol{\xi}=(\xi_1,\xi_2)^\top$ is defined in \eqref{def-g1}-\eqref{def-g2}.
	
	Assume that:
	\begin{itemize}
		\item[(i)] $\mathbb{E}[h_{kj}^2(X)]<\infty$ for all $k=1,\ldots,5$ and $j=1,2$, where the functions $h_{kj}$ are defined in \eqref{defab2}.;
		\item[(ii)] the mapping $\boldsymbol{\xi}$ is continuously differentiable in a neighborhood of $\mathbb{E}[\boldsymbol X]$;
		\item[(iii)] the Jacobian matrix
		\[
		\boldsymbol{A}
		=
		\left.
		\frac{\partial (\xi_1,\xi_2)}{\partial \boldsymbol x} \,
		\right|_{\boldsymbol x=\mathbb{E}[\boldsymbol X]}
		\]
		has full row rank;
		\item[(iv)] 
%
suppose that for some $p\in[a,b]$, the  partial derivative $\partial f_{Y_j}(y;\boldsymbol{\theta},p)/\partial p$
exists on $\mathbb{R}\times(0,\infty)^2\times [a,b]$, and that there exist an integrable function $H_j$ on $D_j$ such that
\begin{align*}
	\left\vert {\partial f_{Y_j}(y;\boldsymbol{\theta},p)\over \partial p}\right\vert
	\leqslant 
	H_j(y), \quad j=1,2.
\end{align*}
	\end{itemize}
	
	Then,
		\begin{align*}
				\begin{pmatrix}
				\widehat{\alpha}\\[0,2cm]
				\widehat{\beta}
			\end{pmatrix}
		\stackrel{\rm a.s.}{\longrightarrow}
			\begin{pmatrix}
				\alpha\\[0,2cm]
				\beta
			\end{pmatrix},
	\end{align*}
	and
	\[
	\sqrt{n}
	\left\{
	\begin{pmatrix}
		\widehat{\alpha}\\[0,2cm]
		\widehat{\beta}
	\end{pmatrix}
	-
	\begin{pmatrix}
		\alpha\\[0,2cm]
		\beta
	\end{pmatrix}
	\right\}
	\xrightarrow{\mathscr D}
	N_2\bigl(\boldsymbol{0}, \boldsymbol{A} \boldsymbol\Sigma\boldsymbol{A}^\top\bigr),
	\]
	where $\boldsymbol\Sigma$ is the covariance matrix of $\boldsymbol X$.
\end{theorem}
%
%
\begin{proof}
Under condition (iv), the dominated convergence theorem allows differentiation under the integral sign, so that the order of integration and partial differentiation can be interchanged; that is,
%
	\begin{align*}
		\mathbb{E}\left[{\partial l_j(\boldsymbol{\theta},p)\over \partial p}\right]
		&=
		n
		\mathbb{E}
		\left[{\partial \log(f_{Y_j}(Y_j;\boldsymbol{\theta},p))\over \partial p}\right]
		\\[0,2cm]
		&=
		n\left[\int {\partial f_{Y_j}(y;\boldsymbol{\theta},p)\over \partial p} {\rm d}y\right] 
		=
		n\left[{\partial \over \partial p} \int { f_{Y_j}(y;\boldsymbol{\theta},p)} {\rm d}y\right] 
		=0, 
		\quad j=1,2.
	\end{align*}
	Thus, using the identity above and taking the expectation of both sides of \eqref{s-eq}, we get:
	\begin{align}\label{derivative-p-0}
(\alpha - 1) \,
&\mathbb{E}\left[
\frac{g_j'\bigl(g_j^{-1}(X)\bigr)}{X}
\, g_j^{-1}(X) \log\bigl(g_j^{-1}(X)\bigr)
\right]
-
(\beta - 1) \, 
\mathbb{E}\left[
\frac{g_j'\bigl(g_j^{-1}(X)\bigr)}{1 - X}
\, g_j^{-1}(X) \log\bigl(g_j^{-1}(X)\bigr)
\right]
\nonumber
\\[0,2cm]
&
=
-
1
-
\mathbb{E}\left[
\log\bigl(g_j^{-1}(X)\bigr)
\right]
-
\mathbb{E}\left[
\frac{g_j''\bigl(g_j^{-1}(X)\bigr)}{g_j'\bigl(g_j^{-1}(X)\bigr)}\, g_j^{-1}(X) \log\bigl(g_j^{-1}(X)\bigr)
\right], 
\quad j=1,2.
	\end{align}
	

Moreover, adopting the notation introduced in \eqref{defab2}, Equation \eqref{derivative-p-0} can be rewritten as
	\begin{align}\label{derivative-p-0-1}
		\alpha
		\mathbb{E}\left[
		h_{1j}(X)
		\right]
		-
		\beta 
		\mathbb{E}\left[
		h_{2j}(X)
		\right]
		=
		-1
		-
		\mathbb{E}\left[
		h_{3j}(X)
		\right]
		-
		\mathbb{E}\left[
		h_{4j}(X)
		\right]
		-
		\mathbb{E}\left[
		h_{5j}(X)
		\right], 
		\quad j=1,2.
	\end{align}
	Using system \eqref{derivative-p-0-1}, we find that
	{\small
	\begin{align}\label{fund-eq}
		\alpha
		=
		\dfrac
		{\displaystyle
				{1+	\mathbb{E}\left[
						h_{31}(X)
						\right]+	\mathbb{E}\left[
						h_{41}(X)
						\right]+	\mathbb{E}\left[
						h_{51}(X)
						\right]\over 	\mathbb{E}\left[
						h_{21}(X)
						\right]}\, 	\mathbb{E}\left[
				h_{22}(X)
				\right]
				-
				1-	\mathbb{E}\left[
				h_{32}(X)
				\right]-	\mathbb{E}\left[
				h_{42}(X)
				\right]-	\mathbb{E}\left[
				h_{52}(X)
				\right]
			}
		{\displaystyle
				\mathbb{E}\left[
				h_{12}(X)
				\right]-{	\mathbb{E}\left[
						h_{11}(X)
						\right]	\mathbb{E}\left[
						h_{22}(X)
						\right]\over 	\mathbb{E}\left[
						h_{21}(X)
						\right]}
			}
	\end{align}
	and
	\begin{align}\label{fund-eq-1}
		\beta
		=
		\dfrac{\alpha\mathbb{E}\left[
				h_{11}(X)
				\right]+1+\mathbb{E}\left[
				h_{31}(X)
				\right]+\mathbb{E}\left[
				h_{41}(X)
				\right]+\mathbb{E}\left[
				h_{51}(X)
				\right]}{\mathbb{E}\left[
				h_{21}(X)
				\right]}.
			\end{align}
	}
	
Since, by the definitions of $\xi_1$ and $\xi_2$ in \eqref{def-g1}-\eqref{def-g2},
	{\footnotesize
	\begin{align*}
		\xi_1(\mathbb{E}[\bm X])
		&=
		\dfrac
{\displaystyle
	{1+	\mathbb{E}\left[
		h_{31}(X)
		\right]+	\mathbb{E}\left[
		h_{41}(X)
		\right]+	\mathbb{E}\left[
		h_{51}(X)
		\right]\over 	\mathbb{E}\left[
		h_{21}(X)
		\right]}\, 	\mathbb{E}\left[
	h_{22}(X)
	\right]
	-
	1-	\mathbb{E}\left[
	h_{32}(X)
	\right]-	\mathbb{E}\left[
	h_{42}(X)
	\right]-	\mathbb{E}\left[
	h_{52}(X)
	\right]
}
{\displaystyle
	\mathbb{E}\left[
	h_{12}(X)
	\right]-{	\mathbb{E}\left[
		h_{11}(X)
		\right]	\mathbb{E}\left[
		h_{22}(X)
		\right]\over 	\mathbb{E}\left[
		h_{21}(X)
		\right]}
},
	\end{align*}
} \noindent
	and
	\begin{align*}
		\xi_2(\mathbb{E}[\bm X])
		&=
		\dfrac{\alpha\mathbb{E}\left[
	h_{11}(X)
	\right]+1+\mathbb{E}\left[
	h_{31}(X)
	\right]+\mathbb{E}\left[
	h_{41}(X)
	\right]+\mathbb{E}\left[
	h_{51}(X)
	\right]}{\mathbb{E}\left[
	h_{21}(X)
	\right]},
	\end{align*}	
combining \eqref{fund-eq}-\eqref{fund-eq-1} with the two identities above yields
		\begin{align}\label{main-id1}
		\boldsymbol{\xi}(\mathbb{E}[\boldsymbol X]) = (\alpha,\beta)^\top.
		\end{align}

Substituting the above identity into \eqref{id-2} establishes the asymptotic normality.
Strong consistency follows from combining \eqref{id-1} with \eqref{main-id1}.
This completes the proof.
\end{proof}


Now, we apply Theorem \ref{clt} to establish the strong consistency and a central limit theorem for the new estimators $\widehat{\alpha}$ and $\widehat{\beta}$ proposed in Section \ref{New estimators}. To this end, the conditions required by Theorem \ref{clt} are verified in detail.

\begin{theorem}\label{clt-1}
	{Let $X \sim \mathrm{Beta}(\boldsymbol{\theta})$ with $\alpha>0$ and $\beta>1$, let $r,s>0$, and let $p\in[a,b]\subset(0,\infty)$.}
If $g_1$ and $g_2$ are the functions defined in Section \ref{New estimators}, that is, $g_1:D_1\equiv(0,1)\to(0,1)$ is the identity function and 
$g_2:D_2\equiv(0,1)\to(0,1)$ is defined by
\[
g_2(x)=(1-x^{1/s})^{1/r}, \quad r,s>0,
\]
then conditions (i)-(iv) of Theorem~\ref{clt} are satisfied.
\end{theorem}
\begin{proof}
For the case $g_1(x)=x$,  we have $g_1^{-1}(x)=x$, $g_1'(x)=1$, and $g_1''(x)=0$. Hence, from the definition \eqref{defab2} of $h_{k1}$, $k=1,2,3,4,5$, it follows that
	\begin{align}\label{h-def-1}
		\begin{array}{llll}
		&h_{11}(x)=\log(x),
		&
		h_{21}(x)=\frac{x}{1-x}\log(x),
		&
		h_{31}(x)=0,
		\\[0,5cm] 
		&
		h_{41}(x)=\frac{2x-1}{1-x}\log(x),
		&
		h_{51}(x)=\log(x).
		&
		\end{array}
	\end{align}
	
	For the case $g_2(x)=(1-x^{1/s})^{1/r}$, we have
	$g_2^{-1}(x)=(1-x^r)^s$,
	\begin{align*}
		g_2'\bigl(g_2^{-1}(x)\bigr)
		=
		-\dfrac{(1-x^r)^{1-s}x^{1-r}}{rs},
		\quad 
		\frac{g_2''\bigl(g_2^{-1}(x)\bigr)}{g_2'\bigl(g_2^{-1}(x)\bigr)}
		=
		\dfrac{(1-x^r)^{-s}x^{-r}\,[r-1+(1-rs)x^r]}{(rs)}.
	\end{align*}
	Consequently, applying \eqref{defab2}, we obtain
	\begin{align}\label{h-def-2}
		\begin{array}{lll}
			&h_{12}(x)
			=
			-\frac{1}{r}\frac{1-x^r}{x^r}\log(1-x^r),
			& 
			h_{22}(x)
			=
			-\frac{1}{r}\frac{1-x^r}{(1-x)x^{r-1}}\log(1-x^r),
			\\[0,5cm] 
			&h_{32}(x)
			=
			\frac{r-1+(1-rs)x^r}{r}
			\frac{1}{x^r}\log(1-x^r),
			&
			h_{42}(x)
			=
			-\frac{2x-1}{r}\frac{1-x^r}{(1-x)x^r}\log(1-x^r),\quad
			\\[0,5cm] 
			&h_{52}(x)
			=
			s\log(1-x^r).
			&
			\end{array}
	\end{align}

{\bf Proof of condition (i)} It is simple to verify that
{\small
\begin{align*}
&\mathbb{E}[\log^2(X)]
=
\psi'(\alpha)-\psi'(\alpha+\beta)
+
[\psi(\alpha)-\psi(\alpha+\beta)]^2;
\\[0,2cm]
&\mathbb{E}\!\left[\frac{X}{1-X}\log^2(X)\right]
=
\frac{B(\alpha+1,\beta-1)}{B(\alpha,\beta)}
\left\{
\psi'(\alpha+1)-\psi'(\alpha+\beta)
+
[\psi(\alpha+1)-\psi(\alpha+\beta)]^2
\right\}, \quad \beta>1;
\\[0,2cm]
&\mathbb{E}\!\left[\left(\frac{X}{1-X}\log X\right)^2\right]
=
\frac{B(\alpha+2,\beta-2)}{B(\alpha,\beta)}
\left\{
\psi'(\alpha+2)-\psi'(\alpha+\beta)
+
[\psi(\alpha+2)-\psi(\alpha+\beta)]^2
\right\}, \quad \beta>2;
\\[0,2cm]
&\mathbb{E}\!\left[\left(\frac{2X-1}{1-X}\log (X)\right)^2\right]
\\[-0,1cm]
&\quad=
4\mathbb{E}\!\left[\left(\frac{X}{1-X}\log (X)\right)^2\right]
-
4\mathbb{E}\!\left[\frac{X}{1-X}\log^2(X)\right]
\\[-0,1cm]
&\qquad
+
\mathbb{E}[\log^2(X)], \quad \beta>2.
\end{align*}
}
{The beta-function representations displayed above are directly valid under their stated sufficient restrictions. The required moments, however, are finite for every $\alpha,\beta>0$: as $x\uparrow1$, $\log x\sim-(1-x)$ cancels the apparent factors $(1-x)^{-1}$, while logarithmic powers are integrable against $x^{\alpha-1}$ as $x\downarrow0$. Hence $\mathbb{E}[h_{k1}^2(X)]<\infty$ for all $k=1,\ldots,5$.}
	
	\medskip
	
Moreover, observe that, for $k=1,\ldots,5$, each expectation $\mathbb{E}[h_{k2}^2(X)]$ can be expressed in terms of the following quantities:
\begin{align*}
	\mathbb{E}\!\left[\log^2(1-X^r)\right], \quad
	\mathbb{E}\!\left[\left(\frac{1-X^r}{X^r}\right)^2 \log^2(1-X^r)\right], \quad
	\mathbb{E}\!\left[\left(\frac{X}{1-X}\right)^2
	\left(\frac{1-X^r}{X^r}\right)^2
	\log^2(1-X^r)\right].
\end{align*}
These expectations are finite under suitable conditions (see Lemmas \ref{lemma-in}, \ref{lemma-in-1}, and \ref{lemma-in-2} in the Appendix). Consequently, $\mathbb{E}[h_{k2}^2(X)] < \infty$ for all $k=1,\ldots,5$.

Therefore, condition (i) of Theorem \ref{clt} holds.

\bigskip 

{\bf Proof of condition (ii)} Note that $\xi_1$ and $\xi_2$, defined in \eqref{def-g1}-\eqref{def-g2}, are rational functions of $\boldsymbol{x}$, hence continuously differentiable on the set
\begin{align*}
	D=\left\{\boldsymbol{x}: x_{21} \neq 0
	\quad \text{and} \quad
	x_{12} - \frac{x_{11}x_{22}}{x_{21}} \neq 0\right\},
\end{align*}
where their denominators do not vanish. In particular, if $\mathbb{E}[\boldsymbol{X}]\in D$, that is,
\begin{align}\label{claim}
\mathbb{E}[h_{21}(X)] \neq 0
\quad \text{and} \quad
\mathbb{E}[h_{12}(X)] - \frac{\mathbb{E}[h_{11}(X)]\mathbb{E}[h_{22}(X)]}{\mathbb{E}[h_{21}(X)]} \neq 0,
\end{align}
then $\boldsymbol{\xi}$ is continuously differentiable in a neighborhood of $\mathbb{E}[\boldsymbol{X}]$.

Therefore, to complete the proof, it suffices to establish \eqref{claim}.
Indeed, the statement $\mathbb{E}[h_{21}(X)] \neq 0$ is immediate, since it is well known that
\begin{align*}
\mathbb{E}[h_{21}(X)]
=
\mathbb{E}\left[\frac{X}{1-X}\log(X)\right]
=
\frac{\alpha}{\beta-1}\left[\psi(\alpha+1)-\psi(\alpha+\beta)\right]
<0, \quad \beta>1.
\end{align*}

On the other hand, we claim that
\begin{align}\label{claim-0}
	{\displaystyle
	\mathbb{E}[h_{12}(X)]-{\mathbb{E}[h_{11}(X)]\mathbb{E}[h_{22}(X)]\over \mathbb{E}[h_{21}(X)]}
	}<0.
\end{align}
Indeed, the above inequality is equivalently to 
\begin{align*}
	{\displaystyle
{\mathbb{E}[h_{22}(X)]}
\over \mathbb{E}[h_{12}(X)]
}
>
{\mathbb{E}[h_{21}(X)]\over \mathbb{E}[h_{11}(X)]},
\end{align*}
because $\mathbb{E}[h_{21}(X)]<0$, $\mathbb{E}[h_{12}(X)]>0$ and $\mathbb{E}[h_{11}(X)]<0$.
Using the definitions of $h_{kj}$ in \eqref{h-def-1} and \eqref{h-def-2}, the above inequality becomes
\begin{align}\label{def-m-r}
	m(r)
	\equiv 
	{\displaystyle
	{\mathbb{E}\left[\frac{X}{1-X}\left\{-\frac{1}{r}\frac{1-X^r}{X^{r}}\log(1-X^r)\right\}\right]}
		\over \displaystyle
	\mathbb{E}\left[-\frac{1}{r}\frac{1-X^r}{X^r}\log(1-X^r)\right]
	}
	>
	{\displaystyle
	\mathbb{E}\left[\frac{X}{1-X}\log(X)\right]
	\over 
	\displaystyle
	\mathbb{E}[\log(X)]},
\end{align}

Note that $m(r)$ can be written as
\begin{align*}
	m(r)
=
	{\displaystyle
		{\mathbb{E}\left[\frac{X}{1-X}\left\{\frac{X^r-1}{r} \, {1\over X^{r}}\, {\log(1-X^r)\over\log(r)}\right\}\right]}
		\over \displaystyle
		\mathbb{E}\left[\frac{X^r-1}{r} \, {1\over X^{r}} \, {\log(1-X^r)\over\log(r)}\right]
	}.
\end{align*}
Since
\begin{align*}
	\lim_{r\to 0^+} \frac{X^r-1}{r}=\log(X),
	\quad 
	\lim_{r\to 0^+} {1\over X^{r}}=1,
	\quad 
	\lim_{r\to 0^+} {\log(1-X^r)\over\log(r)}=1,
\end{align*}
and
\[
\mathbb{E}[\log(X)] = \psi(\alpha) - \psi(\alpha+\beta),
\quad 
\mathbb{E}\!\left[\frac{X}{1-X}\log(X)\right]
=
\frac{\alpha}{\beta-1}\,\mathbb{E}[\log(X)]
+
\frac{1}{\beta-1},
\quad \beta>1,
\]
we get
\begin{align*}
	m(0^+)
	&\equiv 
\lim_{r\to 0^+}m(r)
=	
{\displaystyle
	\mathbb{E}\left[\frac{X}{1-X}\log(X)\right]
	\over 
	\displaystyle
	\mathbb{E}[\log(X)]}
	\\[0,2cm]
	&=
	\frac{\alpha}{\beta-1}
	+
	\frac{1}{(\beta-1)\,[\psi(\alpha)-\psi(\alpha+\beta)]}
	>0, \quad \beta>1.
\end{align*}

Using this fact, inequality \eqref{def-m-r} can be expressed as
\begin{align*}
m(r)>m(0^+), \quad \forall r>0.
\end{align*}

Hence, to conclude \eqref{claim-0}, it suffices to show that the function 
\begin{align}\label{claim-3}
r \mapsto m(r) \quad \text{is strictly increasing on} \ (0,\infty).
\end{align}
Indeed, observe that
\begin{align*}
m'(r)
&=
{\mathbb{E}[u(X) v_r'(X)]\over\mathbb{E}[v_r(X)]}
-
{\mathbb{E}[u(X) v_r(X)]\mathbb{E}[v_r'(X)]\over \mathbb{E}^2[v_r(X)]},
\end{align*}
where we adopt the notation 
\begin{align*}
u(x)
&=
{x\over 1-x},
\\
v_r(x)
&\equiv 
h_{12}(x)
=
-\frac{1}{r} \,\frac{1-x^r}{x^r}\log(1-x^r),
\\
{v_r'(x)\over v_r(x)}
&=
-{1\over r}
-
{1\over r}\,{\log(x^r)\over 1-x^r}
\left[
1
+
{x^r\over \log(1-x^r)}
\right].
\end{align*}

{We now introduce the weighted probability density function}
\begin{align*}
{p_r(x)={v_r(x)f_X(x)\over \mathbb{E}[v_r(X)]}, \quad 0<x<1,}
\end{align*}
Then, we can write
\begin{align*}
m'(r)
&=
\mathbb{E}_r\left[u(X)\, {v_r'(X)\over v_r(X)}\right]
-
\mathbb{E}_r\left[u(X)\right]
\mathbb{E}_r\left[{v_r'(X)\over v_r(X)}\right]
\\
&=
\mathrm{Cov}_r\left(u(X),{v_r'(X)\over v_r(X)}\right),
\end{align*}
where $\mathbb{E}_r$ denotes expectation with respect to the density $p_r$, and $\mathrm{Cov}_r$ is its respective covariance. {Here $v_r'$ denotes differentiation with respect to $r$. Writing $t=x^r$, direct differentiation of $\partial_r\log v_r(x)$, together with $-\log(1-t)>t$ for $0<t<1$, shows that it is strictly increasing in $x$. Since $u(x)=x/(1-x)$ is also strictly increasing, Chebyshev's integral inequality gives}
\begin{align*}
m'(r)>0, \quad \forall r>0.
\end{align*}
That is, the function $r \mapsto m(r)$ is strictly increasing on $(0,\infty)$, which establishes the validity of \eqref{claim-3}. This completes the proof of the second item.

\bigskip 

{\bf Proof of condition (iii)} It suffices to exhibit a nonsingular $2\times 2$ submatrix. A direct computation yields
\[
\frac{\partial(\xi_1,\xi_2)}{\partial(x_{11},x_{12})}
=
\begin{pmatrix}
	\dfrac{\xi_1(\boldsymbol{x}) x_{22}}{x_{21}(x_{12} - \frac{x_{11}x_{22}}{x_{21}})} 
	& 
	-\dfrac{\xi_1(\boldsymbol{x})}{x_{12} - \frac{x_{11}x_{22}}{x_{21}}} \\[12pt]
	\dfrac{\xi_1(\boldsymbol{x})}{x_{21}} 
	+ 
	\dfrac{\xi_1(\boldsymbol{x}) x_{11}x_{22}}{ x_{21}^2(x_{12} - \frac{x_{11}x_{22}}{x_{21}})}
	&
	-\dfrac{\xi_1(\boldsymbol{x}) x_{11}}{x_{21}(x_{12} - \frac{x_{11}x_{22}}{x_{21}})}
\end{pmatrix}.
\]
Its determinant is
\[
\det\left(
\frac{\partial(\xi_1,\xi_2)}{\partial(x_{11},x_{12})}
\right)
=
\frac{\xi_1^2(\boldsymbol{x})}{x_{21}(x_{12}-\frac{x_{11}x_{22}}{x_{21}})}.
\]
By \eqref{claim}, it follows that
\[
\left.
\det\left(
\frac{\partial(\xi_1,\xi_2)}{\partial(x_{11},x_{12})}
\right)
\right|_{\boldsymbol x=\mathbb{E}[\boldsymbol X]}
\neq 0.
\]
Therefore, the above Jacobian is a nonsingular $2\times 2$ submatrix, and consequently
\[
\boldsymbol{A}
=
\left.
\frac{\partial (\xi_1,\xi_2)}{\partial \boldsymbol x}
\right|_{\boldsymbol x=\mathbb{E}[\boldsymbol X]}
\]
has full row rank.

\bigskip 

{\bf Proof of condition (iv)} Using the definition \eqref{A generalized exponential family} of the density $f_{Y_j}(y;\boldsymbol{\theta},p)$, $j=1,2$, with $g_1(x)=x$ and $g_2(x)=(1-x^{1/s})^{1/r}$, we have

\begin{align*}
	{\partial f_{Y_1}(y;\boldsymbol{\theta},p)\over \partial p}
	=
	{\left[
	{1\over p}
	+\alpha\log(y)
	-(\beta-1)
	{y^p\log(y)\over 1-y^p}
	\right]}
	f_{Y_1}(y;\boldsymbol{\theta},p),
\end{align*}

\begin{align*}
	{\partial f_{Y_2}(y;\boldsymbol{\theta},p)\over \partial p}
&=
\Bigg\{
{1\over p}+{1\over s}\log(y)
-{1\over s}\left({\alpha\over r}-1\right)
\frac{y^{p/s}\log(y)}{1-y^{p/s}}
\\[-0,1cm]
&\hspace{2.2cm}
+\frac{\beta-1}{rs}
\frac{y^{p/s}(1-y^{p/s})^{\frac{1}{r}-1}}
{1-(1-y^{p/s})^{1/r}}\log(y)
\Bigg\}
	f_{Y_2}(y;\boldsymbol{\theta},p).
\end{align*}

Hence, for $p\in[a,b]\subset(0,\infty)$, {the endpoint expansions of the two transformed densities give constants $K_1,K_2<\infty$, independent of $p$, such that}
\begin{align*}
 {\left|\frac{\partial f_{Y_1}(y;\boldsymbol{\theta},p)}{\partial p}\right|
 \leqslant K_1(1+|\log y|)\,y^{a\alpha-1}(1-y)^{\beta-1}
 \equiv H_1(y),}
\end{align*}
and
\begin{align*}
 {\left|\frac{\partial f_{Y_2}(y;\boldsymbol{\theta},p)}{\partial p}\right|
 \leqslant K_2(1+|\log y|)\,y^{a\beta/s-1}(1-y)^{\alpha/r-1}
 \equiv H_2(y).}
\end{align*}
To obtain these bounds, use $1-y^p\asymp1-y$ uniformly for
$p\in[a,b]$. For the second transformation,
$1-(1-y^{p/s})^{1/r}\asymp y^{p/s}$ as $y\downarrow0$ and
$1-y^{p/s}\asymp1-y$ as $y\uparrow1$. The score multipliers are bounded
by a constant times $1+|\log y|$. Both $H_1$ and $H_2$ are integrable
because $a\alpha>0$, $a\beta/s>0$, $\beta>0$, and $\alpha/r>0$.
They are genuine dominating functions and do not depend on $p$.
This completes the proof of condition (iv).
\end{proof}

Combining Theorems \ref{clt} and \ref{clt-1} yields the following results.
\begin{corollary}
The new closed-form estimators $\widehat{\alpha}$ and $\widehat{\beta}$ proposed in Section \ref{New estimators} are strongly consistent and satisfy the central limit theorem. In particular, $(\widehat{\alpha},\widehat{\beta})^\top$ is a $\sqrt{n}$-consistent estimator of the true parameters.
\end{corollary}

\begin{corollary}\label{cor:grid-selection}
Let $\mathcal R$ be a fixed finite grid and suppose that the conclusions above
hold for every $r\in\mathcal R$. Then the likelihood-selected estimator in
\eqref{proposed_estimator} is consistent and root-$n$ consistent. Indeed,
the selected error is one of finitely many $O_p(n^{-1/2})$ errors. A single
normal limit for the selected closed-form estimator additionally requires the
selection probability to concentrate on one limiting grid point; it does not
follow from finiteness of the grid alone.
\end{corollary}

\section{Proposing new asymptotically efficient estimators for $\alpha$ and $\beta$}\label{sec:asymptotically_efficient}

Let $\{X_i : i = 1,\ldots,n\}$ be a random sample from 
$X \sim \mathrm{Beta}(\boldsymbol{\theta})$, \(\boldsymbol{\theta}=(\alpha,\beta)^\top\).
Let \(\widehat{\boldsymbol{\theta}}=(\widehat{\alpha},\widehat{\beta})^\top\) 
be the \(\sqrt{n}\)-consistent estimator of
\(\boldsymbol{\theta}\), given in Section \ref{New estimators}.

Starting from the likelihood-selected closed-form estimator
$\widehat{\boldsymbol{\theta}}$ in \eqref{proposed_estimator}, we define the
one-step estimator by one Fisher-scoring update:
\begin{equation}\label{one_step_estimator}
\widetilde{\boldsymbol{\theta}}
=
\widehat{\boldsymbol{\theta}}
+
\frac{1}{n}\,
{I}^{-1}_1(\widehat{\boldsymbol{\theta}})
\nabla \ell_n(\widehat{\boldsymbol{\theta}}),
\end{equation}
where \(\nabla \ell_n(\boldsymbol{\theta})\) denotes the score vector associated with
the log-likelihood function
\[
\ell_n(\boldsymbol{\theta})
=
-n\log(B(\alpha,\beta))
+
(\alpha-1)\sum_{i=1}^n \log(X_i)
+
(\beta-1)\sum_{i=1}^n \log(1-X_i).
\]
Thus, \eqref{proposed_estimator} specifies the closed-form starting estimator,
whereas \eqref{one_step_estimator} specifies its asymptotically efficient
one-step refinement.

The (random) score vector is given by
\begin{align*}
	\nabla\ell_n(\boldsymbol{\theta})
=
n
\begin{pmatrix}
	-\psi(\alpha)+\psi(\alpha+\beta)
	+\displaystyle {1\over n}\sum_{i=1}^n \log(X_i) 
	\\[9pt] \displaystyle
	-\psi(\beta)+\psi(\alpha+\beta)
	+{1\over n}\displaystyle\sum_{i=1}^n \log(1-(X_i))
\end{pmatrix}.
\end{align*}
Moreover, the Fisher information matrix for one observation
is given by
\[
{I}_1(\boldsymbol{\theta})
=
-\,\mathbb{E}_{\boldsymbol{\theta}}
\!\left[\nabla^2\ell_1(\boldsymbol{\theta})\right]
=
\begin{pmatrix}
	\psi'(\alpha)-\psi'(\alpha+\beta) & -\psi'(\alpha+\beta) \\[5pt]
	-\psi'(\alpha+\beta) & \psi'(\beta)-\psi'(\alpha+\beta)
\end{pmatrix},
\]
where $\psi'(x)$ denotes the derivative of the digamma function.

Under the following regularity conditions:
\begin{enumerate}
	\item The distributions $F(\cdot;\boldsymbol{\theta})$ of $X_i$, $i=1,\ldots,n$, share a common support;
	\item The family $\{F(\cdot;\boldsymbol{\theta}) : \boldsymbol{\theta}\in\Theta\}$ is identifiable;
	\item The observations $X_1,\ldots,X_n$ constitute a random sample of size $n$ from the density $f(\cdot;\boldsymbol{\theta})$;
	\item There exists an open set $O\subset \Theta$ containing the true parameter $\boldsymbol{\theta}_0$ such that, for almost all $x$, the density $f(x;\boldsymbol{\theta})$ admits all third-order partial derivatives for every $\boldsymbol{\theta}\in O$;
	\item The identities
	\begin{align*}
		\mathbb{E}_{\boldsymbol{\theta}}\!\left[\frac{\partial \ell_1(\boldsymbol{\theta})}{\partial \theta_i}\right]=0,
		\quad
		[ I_1(\boldsymbol{\theta}) ]_{ij}
		=
		-\,\mathbb{E}_{\boldsymbol{\theta}}\!\left[\frac{\partial^2 \ell_1(\boldsymbol{\theta})}{\partial \theta_i \partial \theta_j}\right],
		\quad i,j=1,2,
	\end{align*}
	hold;
	\item The entries $[I_1(\boldsymbol{\theta})]_{ij}$ are finite and the matrix $I_1(\boldsymbol{\theta})$ is positive definite for all $\boldsymbol{\theta}\in O$;
	\item There exist functions $H_{ijk}$ (possibly depending on $\boldsymbol{\theta}_0$) such that
	\begin{align*}
		\left|\frac{\partial^3 \ell_1(\boldsymbol{\theta})}{\partial \theta_i \partial \theta_j \partial \theta_k}\right|
		\le
		H_{ijk}(x), \quad \forall\, \boldsymbol{\theta}\in O,
	\end{align*}
	and $\mathbb{E}_{\boldsymbol{\theta}_0}[H_{ijk}(X)]<\infty$, for $i,j,k=1,2$;
\end{enumerate}
it follows from Theorem 5.3 of Chapter 6 in \cite{Lehmann-Casella1998} that each component $\widetilde{\theta}_i$, $i=1,2$, is asymptotically efficient, that is,
\[
	{\sqrt{n}
(\widetilde{\boldsymbol\theta}-\boldsymbol\theta)
\;\stackrel{\mathscr{D}}{\longrightarrow}\;
N_2\!\left(\boldsymbol0,{I}_1^{-1}(\boldsymbol\theta)\right),}
\]
{whose marginal variances are the diagonal entries of the inverse Fisher information matrix}
{\small
\[
{I}_1^{-1}(\boldsymbol{\theta})
=
\frac{1}{\psi'(\alpha)\psi'(\beta)
	-\psi'(\alpha)\psi'(\alpha+\beta)
	-\psi'(\beta)\psi'(\alpha+\beta)}
\begin{pmatrix}
	\psi'(\beta)-\psi'(\alpha+\beta) & \psi'(\alpha+\beta)\\[5pt]
	\psi'(\alpha+\beta) & \psi'(\alpha)-\psi'(\alpha+\beta)
\end{pmatrix}.
\]
}

In other words,
\begin{align*}
	\sqrt{n}\,(\widetilde{\alpha} - \alpha) \;\stackrel{\mathscr{D}}{\longrightarrow}\; 
	N\left(0,  
	\frac{\psi'(\beta)-\psi'(\alpha+\beta)}{\psi'(\alpha)\psi'(\beta)
		-\psi'(\alpha)\psi'(\alpha+\beta)
		-\psi'(\beta)\psi'(\alpha+\beta)} \right),
\end{align*}
and
\begin{align*}
	\sqrt{n}\,(\widetilde{\beta} - \beta) \;\stackrel{\mathscr{D}}{\longrightarrow}\; 
	N\left(0, 
	\frac{\psi'(\alpha)-\psi'(\alpha+\beta)}{\psi'(\alpha)\psi'(\beta)
		-\psi'(\alpha)\psi'(\alpha+\beta)
		-\psi'(\beta)\psi'(\alpha+\beta)} \right).
\end{align*}

For finite-sample implementation, we apply step halving to the Fisher-scoring
increment whenever a full update would leave $(0,\infty)^2$ or decrease the
beta log-likelihood relative to the closed-form starting value. This safeguard
preserves admissibility and likelihood ascent and is asymptotically inactive
with probability tending to one.

\begin{remark}
Since the Beta distribution belongs to the exponential family, the above regularity conditions 1-7 hold, and the proofs are omitted for brevity.
\end{remark}


\section{Monte Carlo simulation study}\label{sec:sim}

{This section presents a Monte Carlo study of numerical ML, the closed-form estimators of \citet{Chen-Xiao2025} and \citet{Tamae2020}, the proposed likelihood-selected closed-form estimator in \eqref{proposed_estimator}, and its one-step refinement in \eqref{one_step_estimator}. We also document how the closed-form estimator selects $r$ on a prespecified finite grid.}

\subsection{Finite-sample performance}\label{sec:sim:finite_sample}

We assessed five estimators of the Beta$(\alpha,\beta)$ shape parameters: numerical ML; Chen--Xiao \citep[][Example \ref{ex:closedbeta}]{Chen-Xiao2025}; Tamae et al. \citep[][Example \ref{ex:Tamae2020}]{Tamae2020}; the likelihood-selected member of the closed-form $r$-family in Sect.~\ref{New estimators}, defined by \eqref{proposed_estimator}; and the one-step estimator in \eqref{one_step_estimator}, initialized by that selected closed-form estimator.

We considered the following data-generating scenarios:
$$
n\in\{10,20,50,100\},\quad
\alpha\in\{0.5,1,2\},\quad
\beta\in\{0.5,1,2\}.
$$
We generated $B$ independent and identically distributed (i.i.d.) samples $X^{(b)}=(X^{(b)}_1,\ldots,X^{(b)}_n)$ with $X^{(b)}_i\sim\text{Beta}(\alpha,\beta)$ and computed all five estimators $(\widehat\alpha,\widehat\beta)$ on each replication. We set $B=1000$ and used seed 20260713. The grid was $\mathcal R=\{0.1,0.2,\ldots,2.5\}$. Within each replication, all five procedures were computed from the same simulated sample. Numerical ML used safeguarded Newton iterations on the log-parameter scale, and the one-step estimator used the step-halving safeguard described above. All reported procedures produced finite positive estimates in every replication.

Let $\theta\in\{\alpha,\beta\}$ and $\widehat\theta^{(b)}$ be the estimate on replication $b$. We report, for each estimator and scenario,
\begin{align*}
\text{MARE}(\widehat\theta) &= \frac1B \sum_{b=1}^B \Bigl|\widehat\theta^{(b)}/\theta - 1\Bigr|,
\\[2pt]
\text{RMSE}(\widehat\theta) &= \sqrt{\frac1B \sum_{b=1}^B \bigl(\widehat\theta^{(b)}-\theta\bigr)^2},
\\[2pt]
\text{SE}(\widehat\theta) &= \sqrt{\frac1B \sum_{b=1}^B \Bigl(\widehat\theta^{(b)}-\overline{\widehat\theta}\Bigr)^2},
\quad
\overline{\widehat\theta}=\frac1B\sum_{b=1}^B \widehat\theta^{(b)}.
\end{align*}
{Thus MARE is the mean absolute relative estimation error; it is not the absolute value of relative bias. RMSE combines squared bias and variance, and SE is the Monte Carlo standard deviation.}

Tables~\ref{tab:scenario_metrics_a0p5_b0p5}--\ref{tab:scenario_metrics_a2p0_b2p0}
report all nine data-generating scenarios and $n\in\{10,20,50,100\}$. Error
decreases substantially with $n$ for every method. The likelihood-selected
closed-form estimator already tracks numerical ML closely across the reported
settings. Its one-step refinement removes most of the small remaining
differences and is virtually indistinguishable from ML, especially from $n=20$
onward, as expected from a Fisher-scoring update initialized by a root-$n$
consistent estimator. At $n=10$ no method uniformly dominates: differences vary
with $(\alpha,\beta)$ and should not be overinterpreted with 1000 replications.
Chen--Xiao remains a competitive closed-form alternative, while Tamae et al. is
slightly better in some small-sample settings and slightly more dispersed in
others.

\begingroup
\scriptsize
\renewcommand{\arraystretch}{0.82}
\setlength{\tabcolsep}{3pt}
\captionsetup{font=scriptsize}
\begin{table}[!htbp]
\centering
\caption{Monte Carlo summary for scenario $(\alpha,\beta)=(0.5,0.5)$. Reported metrics per sample size $n$ and estimator. MARE is the mean absolute relative error $\big|\hat\theta/\theta-1\big|$.}
\label{tab:scenario_metrics_a0p5_b0p5}
\begin{tabular}{@{} r l r r r r r r @{}}
\toprule
 \multicolumn{1}{c}{$n$} & Estimator & \multicolumn{1}{c}{$\mathrm{MARE}(\hat{\alpha})$} & \multicolumn{1}{c}{$\mathrm{MARE}(\hat{\beta})$} & \multicolumn{1}{c}{$\mathrm{RMSE}(\hat{\alpha})$} & \multicolumn{1}{c}{$\mathrm{RMSE}(\hat{\beta})$} & \multicolumn{1}{c}{$\mathrm{SE}(\hat{\alpha})$} & \multicolumn{1}{c}{$\mathrm{SE}(\hat{\beta})$} \\
\midrule
\rowcolor{gray!10}   10 & ML & 0.5215 & 0.5810 & 0.4245 & 0.5561 & 0.3856 & 0.5147 \\
\rowcolor{gray!10}   10 & Chen--Xiao & 0.5233 & 0.5804 & 0.4248 & 0.5473 & 0.3849 & 0.5041 \\
\rowcolor{gray!10}   10 & Tamae et al. & 0.5149 & 0.5706 & 0.4162 & 0.5359 & 0.3805 & 0.4987 \\
\rowcolor{gray!10}   10 & Selected closed form & 0.5239 & 0.5808 & 0.4254 & 0.5559 & 0.3860 & 0.5140 \\
\rowcolor{gray!10}   10 & Proposed one-step & 0.5218 & 0.5811 & 0.4245 & 0.5560 & 0.3858 & 0.5147 \\
 20 & ML & 0.2907 & 0.2878 & 0.2179 & 0.2160 & 0.2024 & 0.2031 \\
 20 & Chen--Xiao & 0.2918 & 0.2891 & 0.2179 & 0.2166 & 0.2018 & 0.2032 \\
 20 & Tamae et al. & 0.2977 & 0.2880 & 0.2209 & 0.2154 & 0.2069 & 0.2044 \\
 20 & Selected closed form & 0.2917 & 0.2878 & 0.2182 & 0.2161 & 0.2023 & 0.2029 \\
 20 & Proposed one-step & 0.2908 & 0.2879 & 0.2179 & 0.2160 & 0.2025 & 0.2031 \\
\rowcolor{gray!10}   50 & ML & 0.1635 & 0.1564 & 0.1107 & 0.1034 & 0.1077 & 0.1011 \\
\rowcolor{gray!10}   50 & Chen--Xiao & 0.1640 & 0.1568 & 0.1112 & 0.1035 & 0.1079 & 0.1011 \\
\rowcolor{gray!10}   50 & Tamae et al. & 0.1656 & 0.1609 & 0.1113 & 0.1070 & 0.1089 & 0.1047 \\
\rowcolor{gray!10}   50 & Selected closed form & 0.1642 & 0.1564 & 0.1112 & 0.1034 & 0.1080 & 0.1010 \\
\rowcolor{gray!10}   50 & Proposed one-step & 0.1636 & 0.1564 & 0.1107 & 0.1034 & 0.1077 & 0.1011 \\
100 & ML & 0.1021 & 0.1062 & 0.0672 & 0.0682 & 0.0658 & 0.0668 \\
100 & Chen--Xiao & 0.1021 & 0.1062 & 0.0673 & 0.0683 & 0.0659 & 0.0669 \\
100 & Tamae et al. & 0.1058 & 0.1098 & 0.0695 & 0.0701 & 0.0681 & 0.0690 \\
100 & Selected closed form & 0.1022 & 0.1063 & 0.0673 & 0.0682 & 0.0660 & 0.0668 \\
100 & Proposed one-step & 0.1021 & 0.1062 & 0.0672 & 0.0682 & 0.0658 & 0.0668 \\
\bottomrule
\end{tabular}
\end{table}

\begin{table}[!htbp]
\centering
\caption{Monte Carlo summary for scenario $(\alpha,\beta)=(0.5,1.0)$. Reported metrics per sample size $n$ and estimator. MARE is the mean absolute relative error $\big|\hat\theta/\theta-1\big|$.}
\label{tab:scenario_metrics_a0p5_b1p0}
\begin{tabular}{@{} r l r r r r r r @{}}
\toprule
 \multicolumn{1}{c}{$n$} & Estimator & \multicolumn{1}{c}{$\mathrm{MARE}(\hat{\alpha})$} & \multicolumn{1}{c}{$\mathrm{MARE}(\hat{\beta})$} & \multicolumn{1}{c}{$\mathrm{RMSE}(\hat{\alpha})$} & \multicolumn{1}{c}{$\mathrm{RMSE}(\hat{\beta})$} & \multicolumn{1}{c}{$\mathrm{SE}(\hat{\alpha})$} & \multicolumn{1}{c}{$\mathrm{SE}(\hat{\beta})$} \\
\midrule
\rowcolor{gray!10}   10 & ML & 0.4531 & 0.6437 & 0.3454 & 1.1510 & 0.3112 & 1.0501 \\
\rowcolor{gray!10}   10 & Chen--Xiao & 0.4520 & 0.6439 & 0.3439 & 1.1540 & 0.3092 & 1.0516 \\
\rowcolor{gray!10}   10 & Tamae et al. & 0.4519 & 0.6357 & 0.3400 & 1.1395 & 0.3082 & 1.0499 \\
\rowcolor{gray!10}   10 & Selected closed form & 0.4522 & 0.6422 & 0.3446 & 1.1490 & 0.3104 & 1.0480 \\
\rowcolor{gray!10}   10 & Proposed one-step & 0.4531 & 0.6435 & 0.3453 & 1.1504 & 0.3112 & 1.0496 \\
 20 & ML & 0.2578 & 0.3178 & 0.1826 & 0.4715 & 0.1720 & 0.4402 \\
 20 & Chen--Xiao & 0.2578 & 0.3186 & 0.1824 & 0.4713 & 0.1716 & 0.4390 \\
 20 & Tamae et al. & 0.2603 & 0.3160 & 0.1826 & 0.4636 & 0.1729 & 0.4370 \\
 20 & Selected closed form & 0.2577 & 0.3174 & 0.1824 & 0.4714 & 0.1717 & 0.4399 \\
 20 & Proposed one-step & 0.2578 & 0.3178 & 0.1826 & 0.4714 & 0.1720 & 0.4401 \\
\rowcolor{gray!10}   50 & ML & 0.1463 & 0.1771 & 0.0979 & 0.2422 & 0.0937 & 0.2314 \\
\rowcolor{gray!10}   50 & Chen--Xiao & 0.1462 & 0.1771 & 0.0978 & 0.2422 & 0.0936 & 0.2311 \\
\rowcolor{gray!10}   50 & Tamae et al. & 0.1498 & 0.1788 & 0.0992 & 0.2440 & 0.0954 & 0.2345 \\
\rowcolor{gray!10}   50 & Selected closed form & 0.1461 & 0.1770 & 0.0978 & 0.2422 & 0.0936 & 0.2313 \\
\rowcolor{gray!10}   50 & Proposed one-step & 0.1463 & 0.1771 & 0.0979 & 0.2422 & 0.0937 & 0.2314 \\
100 & ML & 0.0990 & 0.1133 & 0.0644 & 0.1474 & 0.0634 & 0.1454 \\
100 & Chen--Xiao & 0.0990 & 0.1135 & 0.0644 & 0.1477 & 0.0634 & 0.1456 \\
100 & Tamae et al. & 0.1013 & 0.1151 & 0.0655 & 0.1489 & 0.0645 & 0.1470 \\
100 & Selected closed form & 0.0990 & 0.1133 & 0.0644 & 0.1475 & 0.0634 & 0.1454 \\
100 & Proposed one-step & 0.0990 & 0.1133 & 0.0644 & 0.1474 & 0.0634 & 0.1454 \\
\bottomrule
\end{tabular}
\end{table}

\begin{table}[!htbp]
\centering
\caption{Monte Carlo summary for scenario $(\alpha,\beta)=(0.5,2.0)$. Reported metrics per sample size $n$ and estimator. MARE is the mean absolute relative error $\big|\hat\theta/\theta-1\big|$.}
\label{tab:scenario_metrics_a0p5_b2p0}
\begin{tabular}{@{} r l r r r r r r @{}}
\toprule
 \multicolumn{1}{c}{$n$} & Estimator & \multicolumn{1}{c}{$\mathrm{MARE}(\hat{\alpha})$} & \multicolumn{1}{c}{$\mathrm{MARE}(\hat{\beta})$} & \multicolumn{1}{c}{$\mathrm{RMSE}(\hat{\alpha})$} & \multicolumn{1}{c}{$\mathrm{RMSE}(\hat{\beta})$} & \multicolumn{1}{c}{$\mathrm{SE}(\hat{\alpha})$} & \multicolumn{1}{c}{$\mathrm{SE}(\hat{\beta})$} \\
\midrule
\rowcolor{gray!10}   10 & ML & 0.4141 & 0.6753 & 0.3271 & 3.1117 & 0.2990 & 2.9485 \\
\rowcolor{gray!10}   10 & Chen--Xiao & 0.4171 & 0.6789 & 0.3286 & 3.1816 & 0.2999 & 3.0184 \\
\rowcolor{gray!10}   10 & Tamae et al. & 0.4217 & 0.6725 & 0.3293 & 3.1726 & 0.3018 & 3.0257 \\
\rowcolor{gray!10}   10 & Selected closed form & 0.4147 & 0.6749 & 0.3271 & 3.1305 & 0.2991 & 2.9688 \\
\rowcolor{gray!10}   10 & Proposed one-step & 0.4140 & 0.6748 & 0.3270 & 3.1102 & 0.2990 & 2.9474 \\
 20 & ML & 0.2582 & 0.3567 & 0.1863 & 1.0330 & 0.1751 & 0.9427 \\
 20 & Chen--Xiao & 0.2604 & 0.3583 & 0.1878 & 1.0426 & 0.1762 & 0.9507 \\
 20 & Tamae et al. & 0.2639 & 0.3542 & 0.1894 & 1.0306 & 0.1788 & 0.9503 \\
 20 & Selected closed form & 0.2587 & 0.3561 & 0.1865 & 1.0311 & 0.1753 & 0.9413 \\
 20 & Proposed one-step & 0.2581 & 0.3565 & 0.1863 & 1.0325 & 0.1750 & 0.9423 \\
\rowcolor{gray!10}   50 & ML & 0.1388 & 0.1978 & 0.0911 & 0.5406 & 0.0892 & 0.5188 \\
\rowcolor{gray!10}   50 & Chen--Xiao & 0.1399 & 0.1988 & 0.0918 & 0.5442 & 0.0897 & 0.5218 \\
\rowcolor{gray!10}   50 & Tamae et al. & 0.1435 & 0.1986 & 0.0937 & 0.5451 & 0.0917 & 0.5252 \\
\rowcolor{gray!10}   50 & Selected closed form & 0.1391 & 0.1976 & 0.0913 & 0.5403 & 0.0893 & 0.5187 \\
\rowcolor{gray!10}   50 & Proposed one-step & 0.1388 & 0.1977 & 0.0911 & 0.5404 & 0.0892 & 0.5187 \\
100 & ML & 0.0985 & 0.1307 & 0.0628 & 0.3382 & 0.0621 & 0.3321 \\
100 & Chen--Xiao & 0.0990 & 0.1312 & 0.0631 & 0.3389 & 0.0624 & 0.3328 \\
100 & Tamae et al. & 0.1007 & 0.1315 & 0.0644 & 0.3395 & 0.0638 & 0.3341 \\
100 & Selected closed form & 0.0986 & 0.1306 & 0.0629 & 0.3381 & 0.0622 & 0.3322 \\
100 & Proposed one-step & 0.0985 & 0.1307 & 0.0628 & 0.3381 & 0.0621 & 0.3321 \\
\bottomrule
\end{tabular}
\end{table}

\begin{table}[!htbp]
\centering
\caption{Monte Carlo summary for scenario $(\alpha,\beta)=(1.0,0.5)$. Reported metrics per sample size $n$ and estimator. MARE is the mean absolute relative error $\big|\hat\theta/\theta-1\big|$.}
\label{tab:scenario_metrics_a1p0_b0p5}
\begin{tabular}{@{} r l r r r r r r @{}}
\toprule
 \multicolumn{1}{c}{$n$} & Estimator & \multicolumn{1}{c}{$\mathrm{MARE}(\hat{\alpha})$} & \multicolumn{1}{c}{$\mathrm{MARE}(\hat{\beta})$} & \multicolumn{1}{c}{$\mathrm{RMSE}(\hat{\alpha})$} & \multicolumn{1}{c}{$\mathrm{RMSE}(\hat{\beta})$} & \multicolumn{1}{c}{$\mathrm{SE}(\hat{\alpha})$} & \multicolumn{1}{c}{$\mathrm{SE}(\hat{\beta})$} \\
\midrule
\rowcolor{gray!10}   10 & ML & 0.6387 & 0.4565 & 1.1374 & 0.3617 & 1.0343 & 0.3263 \\
\rowcolor{gray!10}   10 & Chen--Xiao & 0.6390 & 0.4555 & 1.1378 & 0.3610 & 1.0330 & 0.3252 \\
\rowcolor{gray!10}   10 & Tamae et al. & 0.6255 & 0.4540 & 1.1188 & 0.3572 & 1.0280 & 0.3244 \\
\rowcolor{gray!10}   10 & Selected closed form & 0.6384 & 0.4561 & 1.1359 & 0.3616 & 1.0315 & 0.3261 \\
\rowcolor{gray!10}   10 & Proposed one-step & 0.6388 & 0.4565 & 1.1372 & 0.3617 & 1.0343 & 0.3263 \\
 20 & ML & 0.3024 & 0.2576 & 0.4243 & 0.1789 & 0.3977 & 0.1683 \\
 20 & Chen--Xiao & 0.3019 & 0.2574 & 0.4231 & 0.1789 & 0.3959 & 0.1682 \\
 20 & Tamae et al. & 0.3027 & 0.2585 & 0.4204 & 0.1794 & 0.3983 & 0.1702 \\
 20 & Selected closed form & 0.3021 & 0.2575 & 0.4240 & 0.1788 & 0.3968 & 0.1682 \\
 20 & Proposed one-step & 0.3025 & 0.2577 & 0.4243 & 0.1789 & 0.3979 & 0.1684 \\
\rowcolor{gray!10}   50 & ML & 0.1818 & 0.1445 & 0.2437 & 0.0953 & 0.2366 & 0.0933 \\
\rowcolor{gray!10}   50 & Chen--Xiao & 0.1817 & 0.1445 & 0.2442 & 0.0953 & 0.2371 & 0.0933 \\
\rowcolor{gray!10}   50 & Tamae et al. & 0.1840 & 0.1473 & 0.2466 & 0.0969 & 0.2400 & 0.0952 \\
\rowcolor{gray!10}   50 & Selected closed form & 0.1817 & 0.1445 & 0.2437 & 0.0953 & 0.2366 & 0.0933 \\
\rowcolor{gray!10}   50 & Proposed one-step & 0.1819 & 0.1445 & 0.2438 & 0.0953 & 0.2367 & 0.0934 \\
100 & ML & 0.1195 & 0.1025 & 0.1575 & 0.0657 & 0.1551 & 0.0647 \\
100 & Chen--Xiao & 0.1196 & 0.1026 & 0.1577 & 0.0657 & 0.1552 & 0.0646 \\
100 & Tamae et al. & 0.1214 & 0.1053 & 0.1590 & 0.0668 & 0.1570 & 0.0658 \\
100 & Selected closed form & 0.1194 & 0.1025 & 0.1576 & 0.0657 & 0.1551 & 0.0647 \\
100 & Proposed one-step & 0.1195 & 0.1025 & 0.1575 & 0.0657 & 0.1551 & 0.0647 \\
\bottomrule
\end{tabular}
\end{table}

\begin{table}[!htbp]
\centering
\caption{Monte Carlo summary for scenario $(\alpha,\beta)=(1.0,1.0)$. Reported metrics per sample size $n$ and estimator. MARE is the mean absolute relative error $\big|\hat\theta/\theta-1\big|$.}
\label{tab:scenario_metrics_a1p0_b1p0}
\begin{tabular}{@{} r l r r r r r r @{}}
\toprule
 \multicolumn{1}{c}{$n$} & Estimator & \multicolumn{1}{c}{$\mathrm{MARE}(\hat{\alpha})$} & \multicolumn{1}{c}{$\mathrm{MARE}(\hat{\beta})$} & \multicolumn{1}{c}{$\mathrm{RMSE}(\hat{\alpha})$} & \multicolumn{1}{c}{$\mathrm{RMSE}(\hat{\beta})$} & \multicolumn{1}{c}{$\mathrm{SE}(\hat{\alpha})$} & \multicolumn{1}{c}{$\mathrm{SE}(\hat{\beta})$} \\
\midrule
\rowcolor{gray!10}   10 & ML & 0.5754 & 0.5254 & 1.0272 & 0.8990 & 0.9392 & 0.8201 \\
\rowcolor{gray!10}   10 & Chen--Xiao & 0.5731 & 0.5237 & 1.0240 & 0.8969 & 0.9361 & 0.8182 \\
\rowcolor{gray!10}   10 & Tamae et al. & 0.5672 & 0.5179 & 1.0111 & 0.8867 & 0.9321 & 0.8163 \\
\rowcolor{gray!10}   10 & Selected closed form & 0.5747 & 0.5251 & 1.0266 & 0.8989 & 0.9384 & 0.8199 \\
\rowcolor{gray!10}   10 & Proposed one-step & 0.5754 & 0.5254 & 1.0272 & 0.8989 & 0.9392 & 0.8200 \\
 20 & ML & 0.2934 & 0.2850 & 0.4325 & 0.4107 & 0.4054 & 0.3841 \\
 20 & Chen--Xiao & 0.2928 & 0.2842 & 0.4315 & 0.4101 & 0.4045 & 0.3836 \\
 20 & Tamae et al. & 0.2926 & 0.2812 & 0.4300 & 0.4071 & 0.4064 & 0.3845 \\
 20 & Selected closed form & 0.2933 & 0.2849 & 0.4322 & 0.4106 & 0.4052 & 0.3840 \\
 20 & Proposed one-step & 0.2934 & 0.2850 & 0.4325 & 0.4106 & 0.4054 & 0.3841 \\
\rowcolor{gray!10}   50 & ML & 0.1564 & 0.1614 & 0.2085 & 0.2148 & 0.2020 & 0.2085 \\
\rowcolor{gray!10}   50 & Chen--Xiao & 0.1562 & 0.1612 & 0.2084 & 0.2146 & 0.2020 & 0.2083 \\
\rowcolor{gray!10}   50 & Tamae et al. & 0.1574 & 0.1626 & 0.2094 & 0.2149 & 0.2037 & 0.2095 \\
\rowcolor{gray!10}   50 & Selected closed form & 0.1563 & 0.1614 & 0.2084 & 0.2148 & 0.2020 & 0.2085 \\
\rowcolor{gray!10}   50 & Proposed one-step & 0.1564 & 0.1614 & 0.2085 & 0.2148 & 0.2020 & 0.2085 \\
100 & ML & 0.1105 & 0.1104 & 0.1467 & 0.1421 & 0.1438 & 0.1388 \\
100 & Chen--Xiao & 0.1104 & 0.1103 & 0.1466 & 0.1420 & 0.1437 & 0.1387 \\
100 & Tamae et al. & 0.1100 & 0.1115 & 0.1464 & 0.1430 & 0.1439 & 0.1401 \\
100 & Selected closed form & 0.1105 & 0.1104 & 0.1467 & 0.1421 & 0.1438 & 0.1388 \\
100 & Proposed one-step & 0.1105 & 0.1104 & 0.1467 & 0.1421 & 0.1438 & 0.1388 \\
\bottomrule
\end{tabular}
\end{table}

\begin{table}[!htbp]
\centering
\caption{Monte Carlo summary for scenario $(\alpha,\beta)=(1.0,2.0)$. Reported metrics per sample size $n$ and estimator. MARE is the mean absolute relative error $\big|\hat\theta/\theta-1\big|$.}
\label{tab:scenario_metrics_a1p0_b2p0}
\begin{tabular}{@{} r l r r r r r r @{}}
\toprule
 \multicolumn{1}{c}{$n$} & Estimator & \multicolumn{1}{c}{$\mathrm{MARE}(\hat{\alpha})$} & \multicolumn{1}{c}{$\mathrm{MARE}(\hat{\beta})$} & \multicolumn{1}{c}{$\mathrm{RMSE}(\hat{\alpha})$} & \multicolumn{1}{c}{$\mathrm{RMSE}(\hat{\beta})$} & \multicolumn{1}{c}{$\mathrm{SE}(\hat{\alpha})$} & \multicolumn{1}{c}{$\mathrm{SE}(\hat{\beta})$} \\
\midrule
\rowcolor{gray!10}   10 & ML & 0.5117 & 0.6159 & 0.8465 & 2.2716 & 0.7634 & 2.0738 \\
\rowcolor{gray!10}   10 & Chen--Xiao & 0.5101 & 0.6176 & 0.8469 & 2.3105 & 0.7635 & 2.1141 \\
\rowcolor{gray!10}   10 & Tamae et al. & 0.5065 & 0.6103 & 0.8442 & 2.3045 & 0.7644 & 2.1246 \\
\rowcolor{gray!10}   10 & Selected closed form & 0.5111 & 0.6157 & 0.8463 & 2.2728 & 0.7632 & 2.0755 \\
\rowcolor{gray!10}   10 & Proposed one-step & 0.5116 & 0.6158 & 0.8464 & 2.2713 & 0.7634 & 2.0736 \\
 20 & ML & 0.2799 & 0.3102 & 0.4073 & 0.8897 & 0.3828 & 0.8280 \\
 20 & Chen--Xiao & 0.2793 & 0.3106 & 0.4064 & 0.8886 & 0.3819 & 0.8270 \\
 20 & Tamae et al. & 0.2802 & 0.3099 & 0.4060 & 0.8844 & 0.3838 & 0.8294 \\
 20 & Selected closed form & 0.2796 & 0.3102 & 0.4070 & 0.8898 & 0.3826 & 0.8283 \\
 20 & Proposed one-step & 0.2799 & 0.3102 & 0.4073 & 0.8895 & 0.3828 & 0.8279 \\
\rowcolor{gray!10}   50 & ML & 0.1612 & 0.1791 & 0.2149 & 0.4753 & 0.2071 & 0.4552 \\
\rowcolor{gray!10}   50 & Chen--Xiao & 0.1615 & 0.1794 & 0.2154 & 0.4766 & 0.2076 & 0.4565 \\
\rowcolor{gray!10}   50 & Tamae et al. & 0.1635 & 0.1797 & 0.2183 & 0.4769 & 0.2111 & 0.4586 \\
\rowcolor{gray!10}   50 & Selected closed form & 0.1612 & 0.1791 & 0.2150 & 0.4751 & 0.2072 & 0.4550 \\
\rowcolor{gray!10}   50 & Proposed one-step & 0.1611 & 0.1791 & 0.2149 & 0.4752 & 0.2071 & 0.4552 \\
100 & ML & 0.1050 & 0.1184 & 0.1372 & 0.2993 & 0.1337 & 0.2914 \\
100 & Chen--Xiao & 0.1053 & 0.1186 & 0.1373 & 0.2993 & 0.1338 & 0.2913 \\
100 & Tamae et al. & 0.1069 & 0.1191 & 0.1387 & 0.2998 & 0.1353 & 0.2930 \\
100 & Selected closed form & 0.1051 & 0.1183 & 0.1372 & 0.2994 & 0.1337 & 0.2914 \\
100 & Proposed one-step & 0.1050 & 0.1184 & 0.1372 & 0.2993 & 0.1337 & 0.2914 \\
\bottomrule
\end{tabular}
\end{table}

\begin{table}[!htbp]
\centering
\caption{Monte Carlo summary for scenario $(\alpha,\beta)=(2.0,0.5)$. Reported metrics per sample size $n$ and estimator. MARE is the mean absolute relative error $\big|\hat\theta/\theta-1\big|$.}
\label{tab:scenario_metrics_a2p0_b0p5}
\begin{tabular}{@{} r l r r r r r r @{}}
\toprule
 \multicolumn{1}{c}{$n$} & Estimator & \multicolumn{1}{c}{$\mathrm{MARE}(\hat{\alpha})$} & \multicolumn{1}{c}{$\mathrm{MARE}(\hat{\beta})$} & \multicolumn{1}{c}{$\mathrm{RMSE}(\hat{\alpha})$} & \multicolumn{1}{c}{$\mathrm{RMSE}(\hat{\beta})$} & \multicolumn{1}{c}{$\mathrm{SE}(\hat{\alpha})$} & \multicolumn{1}{c}{$\mathrm{SE}(\hat{\beta})$} \\
\midrule
\rowcolor{gray!10}   10 & ML & 0.7082 & 0.4642 & 2.2542 & 0.3720 & 1.9758 & 0.3372 \\
\rowcolor{gray!10}   10 & Chen--Xiao & 0.7087 & 0.4657 & 2.2445 & 0.3733 & 1.9612 & 0.3378 \\
\rowcolor{gray!10}   10 & Tamae et al. & 0.7015 & 0.4682 & 2.2228 & 0.3742 & 1.9639 & 0.3401 \\
\rowcolor{gray!10}   10 & Selected closed form & 0.7057 & 0.4629 & 2.2398 & 0.3706 & 1.9604 & 0.3359 \\
\rowcolor{gray!10}   10 & Proposed one-step & 0.7076 & 0.4639 & 2.2514 & 0.3719 & 1.9734 & 0.3371 \\
 20 & ML & 0.3758 & 0.2532 & 1.1903 & 0.1738 & 1.1100 & 0.1627 \\
 20 & Chen--Xiao & 0.3800 & 0.2545 & 1.2145 & 0.1756 & 1.1320 & 0.1640 \\
 20 & Tamae et al. & 0.3775 & 0.2574 & 1.2064 & 0.1778 & 1.1329 & 0.1668 \\
 20 & Selected closed form & 0.3769 & 0.2536 & 1.2020 & 0.1743 & 1.1211 & 0.1631 \\
 20 & Proposed one-step & 0.3757 & 0.2532 & 1.1897 & 0.1737 & 1.1095 & 0.1626 \\
\rowcolor{gray!10}   50 & ML & 0.1930 & 0.1394 & 0.5181 & 0.0927 & 0.5032 & 0.0904 \\
\rowcolor{gray!10}   50 & Chen--Xiao & 0.1937 & 0.1405 & 0.5178 & 0.0934 & 0.5026 & 0.0910 \\
\rowcolor{gray!10}   50 & Tamae et al. & 0.1934 & 0.1435 & 0.5182 & 0.0952 & 0.5048 & 0.0931 \\
\rowcolor{gray!10}   50 & Selected closed form & 0.1929 & 0.1394 & 0.5168 & 0.0926 & 0.5018 & 0.0904 \\
\rowcolor{gray!10}   50 & Proposed one-step & 0.1930 & 0.1394 & 0.5181 & 0.0927 & 0.5031 & 0.0904 \\
100 & ML & 0.1355 & 0.0967 & 0.3550 & 0.0625 & 0.3473 & 0.0614 \\
100 & Chen--Xiao & 0.1355 & 0.0969 & 0.3558 & 0.0628 & 0.3481 & 0.0616 \\
100 & Tamae et al. & 0.1358 & 0.0985 & 0.3562 & 0.0640 & 0.3493 & 0.0629 \\
100 & Selected closed form & 0.1355 & 0.0967 & 0.3549 & 0.0625 & 0.3472 & 0.0614 \\
100 & Proposed one-step & 0.1355 & 0.0967 & 0.3550 & 0.0625 & 0.3473 & 0.0614 \\
\bottomrule
\end{tabular}
\end{table}

\begin{table}[!htbp]
\centering
\caption{Monte Carlo summary for scenario $(\alpha,\beta)=(2.0,1.0)$. Reported metrics per sample size $n$ and estimator. MARE is the mean absolute relative error $\big|\hat\theta/\theta-1\big|$.}
\label{tab:scenario_metrics_a2p0_b1p0}
\begin{tabular}{@{} r l r r r r r r @{}}
\toprule
 \multicolumn{1}{c}{$n$} & Estimator & \multicolumn{1}{c}{$\mathrm{MARE}(\hat{\alpha})$} & \multicolumn{1}{c}{$\mathrm{MARE}(\hat{\beta})$} & \multicolumn{1}{c}{$\mathrm{RMSE}(\hat{\alpha})$} & \multicolumn{1}{c}{$\mathrm{RMSE}(\hat{\beta})$} & \multicolumn{1}{c}{$\mathrm{SE}(\hat{\alpha})$} & \multicolumn{1}{c}{$\mathrm{SE}(\hat{\beta})$} \\
\midrule
\rowcolor{gray!10}   10 & ML & 0.5867 & 0.5146 & 1.9662 & 1.0541 & 1.7939 & 0.9918 \\
\rowcolor{gray!10}   10 & Chen--Xiao & 0.5871 & 0.5145 & 1.9688 & 1.0560 & 1.7967 & 0.9939 \\
\rowcolor{gray!10}   10 & Tamae et al. & 0.5814 & 0.5141 & 1.9541 & 1.0532 & 1.7958 & 0.9949 \\
\rowcolor{gray!10}   10 & Selected closed form & 0.5858 & 0.5143 & 1.9637 & 1.0540 & 1.7918 & 0.9919 \\
\rowcolor{gray!10}   10 & Proposed one-step & 0.5867 & 0.5146 & 1.9662 & 1.0541 & 1.7939 & 0.9918 \\
 20 & ML & 0.3400 & 0.3087 & 1.0514 & 0.4470 & 0.9657 & 0.4074 \\
 20 & Chen--Xiao & 0.3393 & 0.3082 & 1.0498 & 0.4464 & 0.9641 & 0.4069 \\
 20 & Tamae et al. & 0.3356 & 0.3075 & 1.0390 & 0.4450 & 0.9604 & 0.4072 \\
 20 & Selected closed form & 0.3396 & 0.3086 & 1.0499 & 0.4468 & 0.9642 & 0.4073 \\
 20 & Proposed one-step & 0.3400 & 0.3087 & 1.0514 & 0.4470 & 0.9656 & 0.4074 \\
\rowcolor{gray!10}   50 & ML & 0.1815 & 0.1561 & 0.4888 & 0.2107 & 0.4666 & 0.2042 \\
\rowcolor{gray!10}   50 & Chen--Xiao & 0.1816 & 0.1564 & 0.4892 & 0.2109 & 0.4668 & 0.2043 \\
\rowcolor{gray!10}   50 & Tamae et al. & 0.1808 & 0.1581 & 0.4872 & 0.2121 & 0.4670 & 0.2059 \\
\rowcolor{gray!10}   50 & Selected closed form & 0.1814 & 0.1560 & 0.4886 & 0.2106 & 0.4664 & 0.2041 \\
\rowcolor{gray!10}   50 & Proposed one-step & 0.1815 & 0.1561 & 0.4888 & 0.2107 & 0.4666 & 0.2042 \\
100 & ML & 0.1171 & 0.1007 & 0.3059 & 0.1286 & 0.3007 & 0.1264 \\
100 & Chen--Xiao & 0.1173 & 0.1012 & 0.3064 & 0.1290 & 0.3011 & 0.1267 \\
100 & Tamae et al. & 0.1184 & 0.1033 & 0.3075 & 0.1311 & 0.3031 & 0.1290 \\
100 & Selected closed form & 0.1171 & 0.1007 & 0.3059 & 0.1287 & 0.3007 & 0.1264 \\
100 & Proposed one-step & 0.1171 & 0.1007 & 0.3059 & 0.1286 & 0.3007 & 0.1264 \\
\bottomrule
\end{tabular}
\end{table}

\begin{table}[!htbp]
\centering
\caption{Monte Carlo summary for scenario $(\alpha,\beta)=(2.0,2.0)$. Reported metrics per sample size $n$ and estimator. MARE is the mean absolute relative error $\big|\hat\theta/\theta-1\big|$.}
\label{tab:scenario_metrics_a2p0_b2p0}
\begin{tabular}{@{} r l r r r r r r @{}}
\toprule
 \multicolumn{1}{c}{$n$} & Estimator & \multicolumn{1}{c}{$\mathrm{MARE}(\hat{\alpha})$} & \multicolumn{1}{c}{$\mathrm{MARE}(\hat{\beta})$} & \multicolumn{1}{c}{$\mathrm{RMSE}(\hat{\alpha})$} & \multicolumn{1}{c}{$\mathrm{RMSE}(\hat{\beta})$} & \multicolumn{1}{c}{$\mathrm{SE}(\hat{\alpha})$} & \multicolumn{1}{c}{$\mathrm{SE}(\hat{\beta})$} \\
\midrule
\rowcolor{gray!10}   10 & ML & 0.5997 & 0.5799 & 2.0375 & 1.9527 & 1.8317 & 1.7579 \\
\rowcolor{gray!10}   10 & Chen--Xiao & 0.5983 & 0.5780 & 2.0324 & 1.9458 & 1.8289 & 1.7533 \\
\rowcolor{gray!10}   10 & Tamae et al. & 0.5925 & 0.5737 & 2.0161 & 1.9299 & 1.8239 & 1.7485 \\
\rowcolor{gray!10}   10 & Selected closed form & 0.5994 & 0.5798 & 2.0368 & 1.9527 & 1.8315 & 1.7579 \\
\rowcolor{gray!10}   10 & Proposed one-step & 0.5997 & 0.5798 & 2.0375 & 1.9527 & 1.8317 & 1.7579 \\
 20 & ML & 0.2873 & 0.2912 & 0.8143 & 0.8222 & 0.7523 & 0.7630 \\
 20 & Chen--Xiao & 0.2866 & 0.2907 & 0.8123 & 0.8204 & 0.7516 & 0.7623 \\
 20 & Tamae et al. & 0.2855 & 0.2904 & 0.8074 & 0.8159 & 0.7525 & 0.7629 \\
 20 & Selected closed form & 0.2872 & 0.2912 & 0.8140 & 0.8222 & 0.7523 & 0.7630 \\
 20 & Proposed one-step & 0.2873 & 0.2912 & 0.8142 & 0.8222 & 0.7523 & 0.7630 \\
\rowcolor{gray!10}   50 & ML & 0.1645 & 0.1709 & 0.4379 & 0.4514 & 0.4196 & 0.4358 \\
\rowcolor{gray!10}   50 & Chen--Xiao & 0.1643 & 0.1706 & 0.4373 & 0.4507 & 0.4193 & 0.4354 \\
\rowcolor{gray!10}   50 & Tamae et al. & 0.1643 & 0.1698 & 0.4371 & 0.4489 & 0.4207 & 0.4352 \\
\rowcolor{gray!10}   50 & Selected closed form & 0.1645 & 0.1709 & 0.4379 & 0.4514 & 0.4196 & 0.4358 \\
\rowcolor{gray!10}   50 & Proposed one-step & 0.1645 & 0.1709 & 0.4379 & 0.4514 & 0.4196 & 0.4358 \\
100 & ML & 0.1166 & 0.1141 & 0.2987 & 0.2939 & 0.2934 & 0.2889 \\
100 & Chen--Xiao & 0.1166 & 0.1140 & 0.2986 & 0.2936 & 0.2934 & 0.2887 \\
100 & Tamae et al. & 0.1175 & 0.1141 & 0.2993 & 0.2934 & 0.2947 & 0.2889 \\
100 & Selected closed form & 0.1167 & 0.1141 & 0.2988 & 0.2939 & 0.2935 & 0.2889 \\
100 & Proposed one-step & 0.1166 & 0.1141 & 0.2987 & 0.2939 & 0.2934 & 0.2889 \\
\bottomrule
\end{tabular}
\end{table}
\endgroup

\clearpage

\subsection{{Likelihood-guided grid selections}}\label{sec:sim:profile}

To describe the likelihood-selected closed-form estimator in
\eqref{proposed_estimator}, which also initializes \eqref{one_step_estimator}, we recorded the
likelihood-selected value $\widehat r$ in every Monte Carlo replication. The grid
was $\mathcal R=\{0.1,0.2,\ldots,2.5\}$. Table~\ref{tab:rs_mode_n100} reports the
modal value at $n=100$, its count over 1000 replications, and the corresponding
percentage.

\begin{table}[!ht]
\centering
\caption{Modal likelihood-selected value of $r$ at $n=100$ over 1000 replications.}
\label{tab:rs_mode_n100}
\begin{tabular}{@{} c c r r @{}} \toprule
$(\alpha,\beta)$ & Modal $\widehat r$ & Count & \% \\ \midrule
(0.5,0.5) & 0.8 & 210 & 21.0\\
(0.5,1.0) & 0.8 & 429 & 42.9\\
(0.5,2.0) & 0.7 & 516 & 51.6\\
(1.0,0.5) & 1.2 & 244 & 24.4\\
(1.0,1.0) & 1.0 & 530 & 53.0\\
(1.0,2.0) & 0.7 & 351 & 35.1\\
(2.0,0.5) & 2.5 & 170 & 17.0\\
(2.0,1.0) & 1.5 & 161 & 16.1\\
(2.0,2.0) & 1.0 & 203 & 20.3\\
\bottomrule
\end{tabular}
\end{table}

Selection is scenario-dependent: the modal value ranges from $0.7$ to $2.5$ and
does not universally concentrate at $r=1$. The frequencies also show that the
modal grid point need not dominate the sampling distribution of $\widehat r$.
This observation is consistent with Corollary~\ref{cor:grid-selection}: finite-grid
selection preserves the root-$n$ rate, but a unique normal limit for the selected
closed-form starting estimator requires an additional concentration condition.

\subsection{Remarks about the simulation results}
Across the scenarios considered, the likelihood-selected closed-form estimator
provides a direct assessment of the proposed $r$-family against the existing
closed-form benchmarks of Chen--Xiao and Tamae et al. Its one-step refinement
remains close to numerical ML and quantifies the gain obtained from a single
Fisher-scoring update. The two proposed procedures therefore play distinct roles:
\eqref{proposed_estimator} retains closed-form parameter estimates followed by a
finite likelihood-guided grid selection, whereas \eqref{one_step_estimator} adds
one controlled numerical step to attain asymptotic efficiency. The grid limits
and spacing are explicit design choices and should not be described as
tuning-free.


\section{Application to real data: Proportion of land used for farming}\label{Sec:application}

\begin{figure}[!ht]
  \centering
  \includegraphics[width=0.9\linewidth]{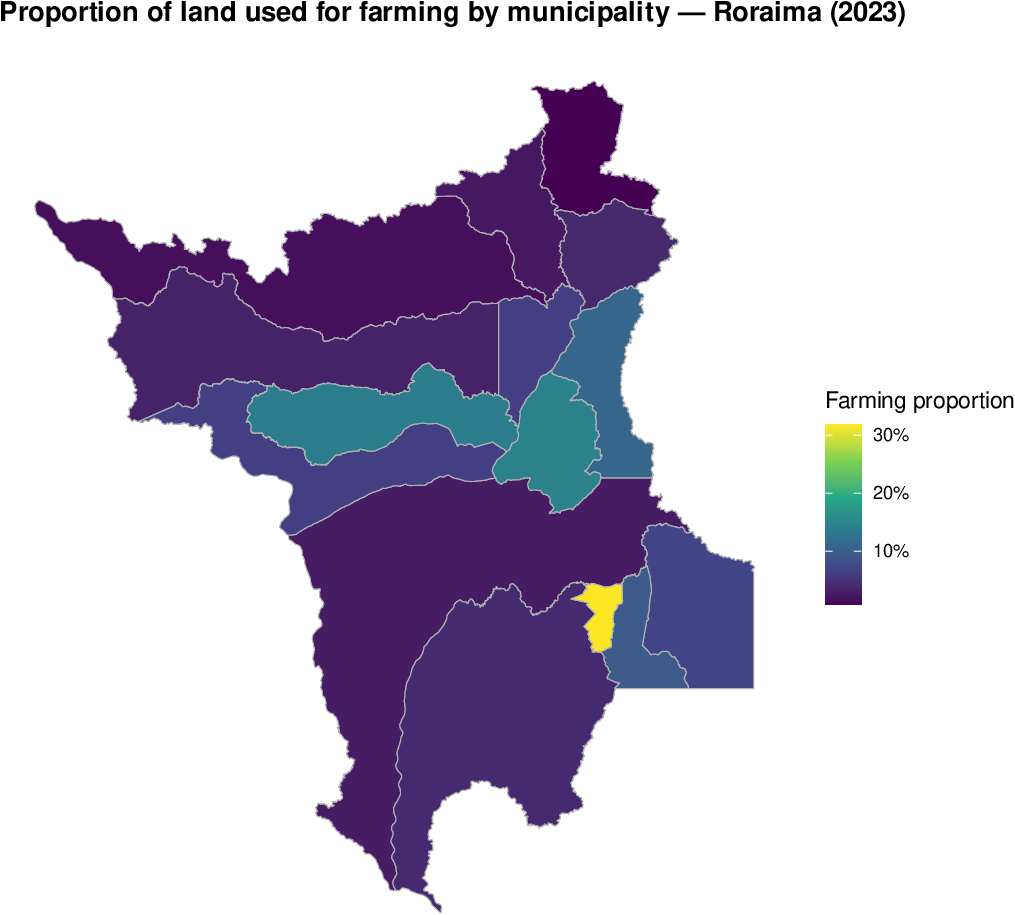}
  \caption{Proportion of land used for farming by municipality - Roraima (2023).}
  \label{fig:map-farming-rr}
\end{figure}

We illustrate the proposed methodology with municipal data from the state of Roraima (Brazil) for 2023. This state is the northernmost one, wedged between Venezuela and Guyana, and almost entirely covered by the Amazon biome. For each municipality $i=1,\ldots,15$, let $X_i\in(0,1)$ denote the proportion of its total area used for farming (crop and pasture). Table~\ref{tab:farming_rr_2023_noyear} reports proportion of land used for farming for each municipality, and Figure~\ref{fig:map-farming-rr} maps the spatial distribution of $X_i$ across the state. The proportion data were computed using area information available at \url{https://brasil.mapbiomas.org/en/}. Most municipalities display relatively small farming shares, with a few markedly larger values (e.g., Cant\'a, Mucaja\'i, S\~ao Luiz), which suggests a positively skewed distribution on $(0,1)$. In this case, the beta distribution is a natural choice for proportion data as its support matches the unit interval and it flexibly accommodates a wide range of shapes.

The 15 observations comprise all municipalities in Roraima, rather than a random
sample of municipalities from a larger population. We therefore treat the beta
fit as a descriptive approximation and use likelihood values to compare numerical
behavior; standard i.i.d. sampling uncertainty is not interpreted as design-based
inference for the state.

\begin{table}[!ht]
\centering
\caption{Proportion of land used for farming by municipality - State of Roraima, Brazil (2023)}
\label{tab:farming_rr_2023_noyear}
\small
\begin{tabular}{l r r r}
\toprule
municipality &  & prop. farming \\
\midrule
Amajari                &  & 0.016976117 \\
Alto Alegre            &  & 0.033823575 \\
Boa Vista              &   & 0.063730010 \\
Bonfim                 &   & 0.110266536 \\
Cantá                  &   & 0.145216265 \\
Caracaraí              &  & 0.026929221 \\
Caroebe                &  & 0.070832843 \\
Iracema                &  & 0.065587585 \\
Mucajaí                & & 0.136861919 \\
Normandia              &   & 0.038840926 \\
Pacaraima              &   & 0.023939695 \\
Rorainópolis           &  & 0.040946891 \\
São João da Baliza     &   & 0.095468970 \\
São Luiz               &   & 0.317332200 \\
Uiramutã               &   & 0.008077883 \\
\bottomrule
\end{tabular}
\end{table}

{We fit the $\operatorname{Beta}(\alpha,\beta)$ model and compare numerical ML, Chen--Xiao, Tamae et al., the likelihood-selected closed-form starting estimator, and the proposed one-step estimator. We used $\mathcal R=\{0.1,0.2,\ldots,2.5\}$ and the admissibility and tie rules in Remark~\ref{remark:profile}.}

Table~\ref{tab:beta-fit-compare} reports the point estimates and beta
log-likelihoods. AIC and BIC are not used to compare these rows because all of them
estimate the same two-parameter beta model; with the same parameter count, those
criteria merely reorder the log-likelihood values. The one-step fit has estimated
mean $\widehat{\mathbb E}[X]=\widehat\alpha/(\widehat\alpha+\widehat\beta)=0.0806$.
The largest observed share, 0.3173322 for S\~ao Luiz, is retained in the analysis.

\begin{table}[!ht]
\centering
\caption{Beta estimates and log-likelihood values for the Roraima farming proportions.}
\label{tab:beta-fit-compare}
\small
\begin{tabular}{lrrrr}
\toprule
Estimator & $\widehat\alpha$ & $\widehat\beta$ & $\ell_n$ & $\widehat r$ \\
\midrule
ML & 1.29168 & 14.73041 & 23.23920 & -- \\
Chen--Xiao & 1.21300 & 13.86324 & 23.22123 & -- \\
Tamae et al. & 1.19376 & 13.79285 & 23.21011 & -- \\
Selected closed form & 1.27584 & 14.79167 & 23.23605 & 0.1 \\
Proposed one-step & 1.29114 & 14.72345 & 23.23920 & 0.1 \\
\bottomrule
\end{tabular}
\end{table}

The selected value $\widehat r=0.1$ is the lower endpoint of the prespecified
grid. As a sensitivity check, we maximized the likelihood along the closed-form
$r$-family over the wider continuous interval $[0.001,2.5]$. The optimum was
$r=0.08574$, with $(\widehat\alpha_r,\widehat\beta_r)=(1.27787,14.82168)$ and
$\ell_n=23.23607$, compared with $\ell_n=23.23605$ at $r=0.1$. These negligible
differences show that the substantive conclusions are insensitive to the lower
grid endpoint. We retain the prespecified finite grid so that the application
uses the same estimator definition as the Monte Carlo study.

\begin{figure}[!ht]
  \centering
  \includegraphics[width=0.6\linewidth]{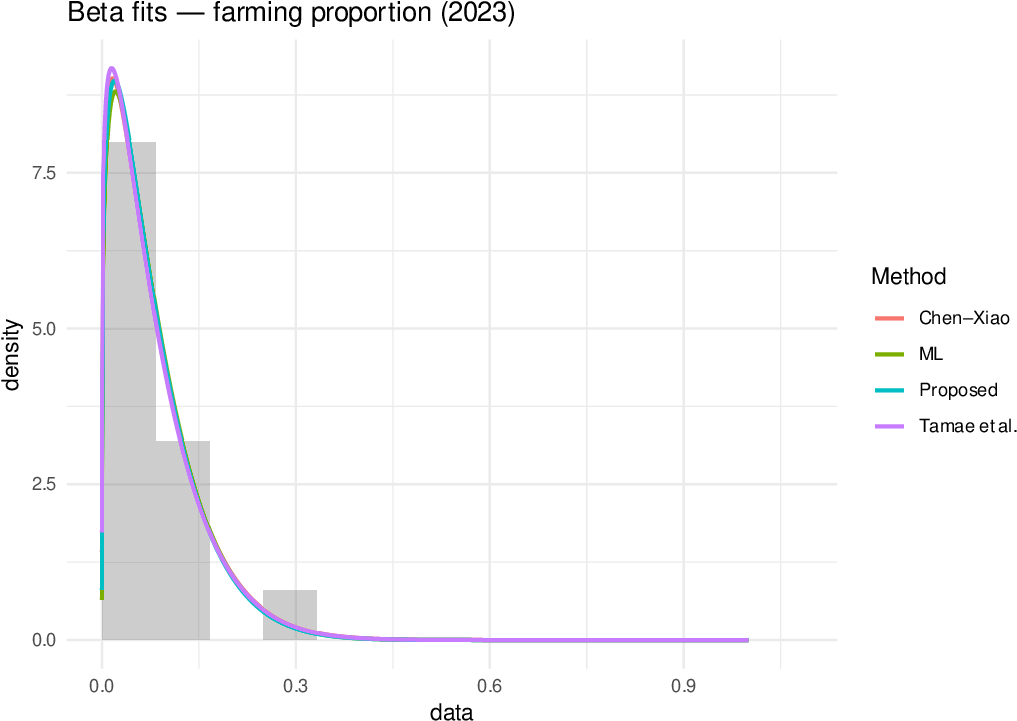}
  \caption{Beta fits based on farming proportion data (2023).}
  \label{fig:beta-fits-farming}
\end{figure}

\begin{figure}[!ht]
  \centering
  \includegraphics[width=0.6\linewidth]{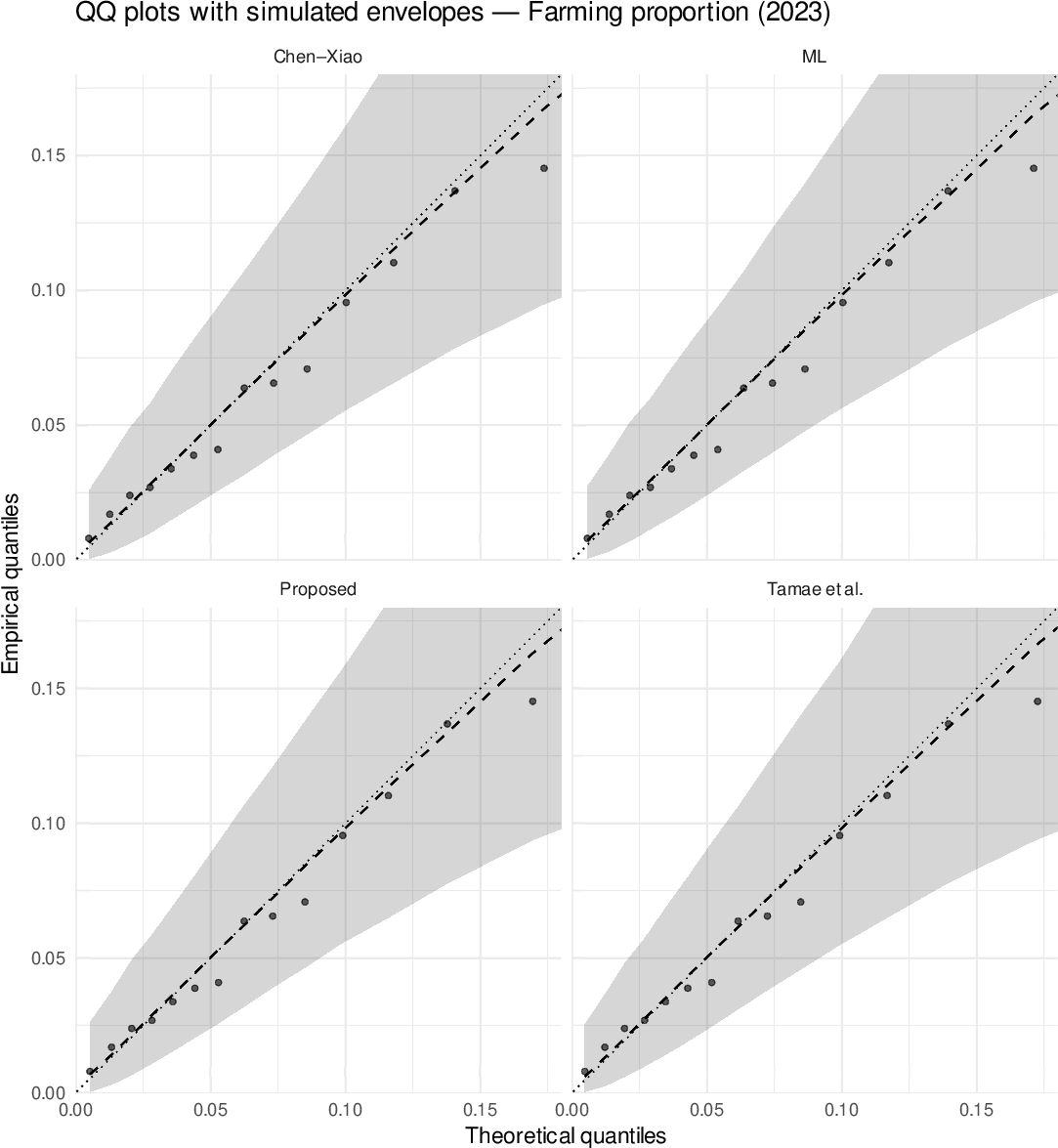}
  \caption{QQ plots with simulated envelopes based on farming proportion data (2023).}
  \label{fig:qq-envelopes}
\end{figure}

Figure~\ref{fig:beta-fits-farming} shows the fitted beta densities superimposed
on the histogram. For legibility, Figures~\ref{fig:beta-fits-farming} and
\ref{fig:qq-envelopes} display four fits; the label ``Proposed'' denotes the
proposed one-step estimator. The selected closed-form fit is omitted from these
plots because it is visually indistinguishable at the displayed scale, but its
numerical results are reported in Table~\ref{tab:beta-fit-compare}. The QQ
envelopes are retained as exploratory diagnostics and are not used to make a
formal goodness-of-fit claim for this small, spatially structured census.

\clearpage

In short, the application illustrates that the closed-form starting value and
its one-step refinement behave similarly to numerical ML for this small, skewed
set of proportions.

\section{Concluding remarks}\label{sec:concluding}

Power transformations provide a direct route from transformed-model likelihood
scores to unbiased estimating equations for the beta shape parameters. The
construction recovers existing closed-form procedures and yields a new family
indexed by $r$; the second transformation parameter $s$ cancels algebraically.
For each fixed admissible $r$, the estimator is consistent and asymptotically
normal. Selection over the reported finite grid preserves root-$n$ consistency,
and one Fisher-scoring update attains the inverse Fisher-information limit.

The Monte Carlo study evaluates both the likelihood-selected closed-form
estimator and its one-step refinement while retaining all nine original
parameter scenarios. This separation distinguishes the finite-sample behavior
of the new closed-form family from the additional gain produced by one
Fisher-scoring update. The Roraima example illustrates both procedures
descriptively on a small, skewed census of municipal proportions. The results
should be understood as beta-specific. Extensions to other families require new
transformations and separate verification of unbiasedness, moment conditions,
identifiability, and nonsingularity of the population equations.


\section*{Statements and Declarations}

\paragraph*{Funding}
This work was supported in part by the Conselho Nacional de Desenvolvimento
Cient\'ifico e Tecnol\'ogico (CNPq; grant 304716/2023-5) and the Coordena\c{c}\~ao
de Aperfei\c{c}oamento de Pessoal de N\'ivel Superior (CAPES), Brazil.

\paragraph*{Competing interests}
The authors have no relevant financial or non-financial interests to disclose.

\paragraph*{Author contributions}
R. Vila: Conceptualization; Methodology; Formal analysis; Software; Validation;
Visualization; Writing -- original draft; Writing -- review \& editing.
H. Saulo: Conceptualization; Methodology; Investigation; Data curation;
Resources; Supervision; Project administration; Funding acquisition;
Validation; Writing -- review \& editing.
F. Quintino: Formal analysis; Validation; Visualization; Writing -- review \&
editing.
P. L. Ramos: Methodology; Software; Validation; Writing -- review \& editing.

\paragraph*{Data availability}
The complete data set analyzed in Section~\ref{Sec:application} is reproduced in
Table~\ref{tab:farming_rr_2023_noyear}; the underlying land-cover platform is
available at \url{https://brasil.mapbiomas.org/en/}.

\paragraph*{Code availability}
The R code used for the Monte Carlo study and the empirical application is
available from the corresponding author upon reasonable request.



\newpage
\begin{appendices}

\section{Finiteness of expectations for functions of a beta random variable}

\noindent

In this section, we present some auxiliary results from Section \ref{sect-8}.

\begin{lemma}\label{lemma-in}
	Let $X \sim \mathrm{Beta}(\alpha,\beta)$ with $\alpha,\beta>0$ and $r>0$. Then
	\[
	\mathbb{E}\left[\log^2(1-X^r)\right] < \infty.
	\]
\end{lemma}

\begin{proof}
	Fix $\varepsilon \in (0,1/2)$, we can write
	\begin{align}
		\mathbb{E}\left[\log^2(1-X^r)\right]
		&=
		\frac{1}{B(\alpha,\beta)}
		\Bigg[
		\int_0^\varepsilon \log^2(1-x^r)\,x^{\alpha-1}(1-x)^{\beta-1}\,{\rm d}x
		\nonumber
		\\[0,2cm]
		&+
		\int_\varepsilon^{1-\varepsilon}
		\log^2(1-x^r)\,x^{\alpha-1}(1-x)^{\beta-1}\,{\rm d}x
		+
		\int_{1-\varepsilon}^1
		\log^2(1-x^r)\,x^{\alpha-1}(1-x)^{\beta-1}\,{\rm d}x
		\Bigg]. \label{int-0}
	\end{align}
	
	On the compact interval $[\varepsilon,1-\varepsilon]$, the integrand is continuous and therefore bounded, so
	\begin{align}\label{int-1}
		\int_\varepsilon^{1-\varepsilon} \log^2(1-x^r)\,x^{\alpha-1}(1-x)^{\beta-1}\,{\rm d}x < \infty.
	\end{align}
	
	It remains to analyze the behavior near the endpoints.

	Using $\log(1-x^r)\sim -x^r$ as $x\to 0^+$, there exists $C>0$ such that
	$
	|\log(1-x^r)| \leqslant C x^r$,
	for $x \in (0,\varepsilon).$
	Hence
	\[
	\log^2(1-x^r) x^{\alpha-1}(1-x)^{\beta-1}
	\leqslant
	C^2 x^{\alpha-1+2r}.
	\]
	Since $\alpha>0$, we have $\alpha-1+2r>-1$, so
	\begin{align*}
		\int_0^\varepsilon \log^2(1-x^r)\,x^{\alpha-1}(1-x)^{\beta-1}\,{\rm d}x
		\leqslant
		C^2
		\int_0^\varepsilon x^{\alpha-1+2r}\,{\rm d}x < \infty.
	\end{align*}

	On the other hand, using $1-x^r \sim r(1-x)$ as $x\to 1^-$, there exists $C'>0$ such that
	$
	|\log(1-x^r)| \leqslant C' |\log(1-x)|
	$
	for $x \in (1-\varepsilon,1).$
	Thus
	\[
	\log^2(1-x^r) x^{\alpha-1}(1-x)^{\beta-1}
	\leqslant
	(C')^2 (1-x)^{\beta-1}\log^2(1-x).
	\]
	It is well known that
	$
	\int_{1-\varepsilon}^1 (1-x)^{\beta-1}\log^2(1-x)\,{\rm d}x < \infty$, for all $\beta>0$. Hence
	\begin{align}\label{int-3}
		\int_{1-\varepsilon}^1
		\log^2(1-x^r)\,x^{\alpha-1}(1-x)^{\beta-1}\,{\rm d}x
		\leqslant
		(C')^2 
		\int_{1-\varepsilon}^1
		(1-x)^{\beta-1}\log^2(1-x)
		\,{\rm d}x< \infty.
	\end{align}
	
	Combining \eqref{int-0}-\eqref{int-3}, the result follows.
\end{proof}

\begin{lemma}\label{lemma-in-1}
	Let $X \sim \mathrm{Beta}(\alpha,\beta)$ with $\alpha,\beta>0$ and $r>0$. Then
	\begin{align}\label{exp-1}
		\mathbb{E}\!\left[\left(\frac{1-X^r}{X^r}\right)^2 \log^2(1-X^r)\right] < \infty,
		\quad {\alpha,\beta>0.}
	\end{align}
\end{lemma}

\begin{proof}
	%
	Let $\varepsilon \in (0,1/2)$ be fixed, and decompose
	the expectation in \eqref{exp-1} into three parts, as in the proof of Lemma \ref{lemma-in}.
	
	We note that on the compact interval $[\varepsilon,1-\varepsilon]$ the integrand
	\[
	\left(\frac{1-x^r}{x^r}\right)^2 \log^2(1-x^r)\,x^{\alpha-1}(1-x)^{\beta-1}
	\]
	is continuous and therefore bounded, which ensures its integrability.
	
	It remains to examine its behavior in neighborhoods of the endpoints.
	
	{Since $\log(1-x^r)\sim-x^r$ as $x\to0^+$, there exists $C>0$ such that $|\log(1-x^r)|\leq Cx^r$ for $x\in(0,\varepsilon)$. Therefore,}
	\[
	\left(\frac{1-x^r}{x^r}\right)^2 \log^2(1-x^r) x^{\alpha-1}(1-x)^{\beta-1}
	\leqslant
	{C x^{\alpha-1}.}
	\]
	Hence
	\[
	\int_0^\varepsilon
	\left(\frac{1-x^r}{x^r}\right)^2 \log^2(1-x^r)\,x^{\alpha-1}(1-x)^{\beta-1}
	\,{\rm d}x
	\leqslant
	C
	\int_0^\varepsilon x^{\alpha-1}\,{\rm d}x < \infty, \quad 
	{\alpha>0.}
	\]
	
	On the other hand,
	since $1-x^r \sim r(1-x)$ as $x\to 1^-$, there exists $C'>0$ such that
	$
	|\log(1-x^r)| \leqslant C' |\log(1-x)|,
	\ \text{for } x \in (1-\varepsilon,1).
	$
	Moreover, $x^r \to 1$, so
	${(1-x^r)}/{x^r} \asymp (1-x)$, that is,
	there exist constants $c_1,c_2>0$ and a neighborhood of $1$ such that
	$
	c_1(1-x) \leqslant {(1-x^r)}/{x^r} \leqslant c_2(1-x),
	$
	for all $x$ sufficiently close to $1$.
	Hence, for some $C''>0$,
	\[
	\left(\frac{1-x^r}{x^r}\right)^2 \log^2(1-x^r)
	\leqslant
	C'' (1-x)^2 \log^2(1-x).
	\]
	Therefore, the integrand is bounded by
	\[
	\int_{1-\varepsilon}^1
	\left(\frac{1-x^r}{x^r}\right)^2 \log^2(1-x^r)\,x^{\alpha-1}(1-x)^{\beta-1}
	\,{\rm d}x
	\leqslant
	C'' 		\int_{1-\varepsilon}^1(1-x)^{\beta+1} \log^2(1-x)	\,{\rm d}x<\infty,
	\quad \beta>0.
	\]
	
	\medskip
	{Combining the three parts, we conclude that the integral is finite for every $\alpha,\beta>0$.}
\end{proof}

\begin{lemma}\label{lemma-in-2}
	Let $X \sim \mathrm{Beta}(\alpha,\beta)$ with $\alpha,\beta>0$ and $r>0$. Then
	\begin{align}\label{exp-2}
		\mathbb{E}\!\left[\left(\frac{X}{1-X}\right)^2
		\left(\frac{1-X^r}{X^r}\right)^2
		\log^2(1-X^r)\right] < \infty.
	\end{align}
\end{lemma}

\begin{proof}
	Let $\varepsilon \in (0,1/2)$ be fixed, and decompose
	the expectation in \eqref{exp-2} into three parts, as in the proof of Lemma \ref{lemma-in}.
	
	%
	On the compact interval $[\varepsilon,1-\varepsilon]$, the integrand is continuous and bounded, hence integrable.
	
	It remains to analyze the behavior near $0$ and $1$.
	
	As $x \to 0^+$, using $\log(1-u)\sim -u$ as $u\to 0$, we obtain
	$
	\log(1-x^r)\sim -x^r,
	\ \text{and hence} \
	\left[{(1-x^r)}/{x^r}\right]^2 \log^2(1-x^r) \sim 1.
	$
	Therefore,
	\[
	\left(\frac{x}{1-x}\right)^2
	\left(\frac{1-x^r}{x^r}\right)^2
	\log^2(1-x^r)
	\sim x^2,
	\]
	and the integrand behaves like $x^{\alpha+1}$ near $0$, which is integrable for all $\alpha>0$.
	
	As $x \to 1^-$, we have $1-x^r \sim r(1-x)$ and $x^r \to 1$, so
	$
	\left[{(1-x^r)}/{x^r}\right]^2 \log^2(1-x^r)
	\sim (1-x)^2 \log^2(1-x).
	$
	Thus,
	\[
	\left(\frac{x}{1-x}\right)^2
	\left(\frac{1-x^r}{x^r}\right)^2
	\log^2(1-x^r)
	\sim \log^2(1-x),
	\]
	and the integrand behaves like
	\[
	(1-x)^{\beta-1}\log^2(1-x),
	\]
	which is integrable near $1$ whenever $\beta>0$.
	
	Combining these arguments, the integral is finite for all $\alpha,\beta>0$.
\end{proof}

\end{appendices}
\end{document}